\documentclass[11pt,preprint]{elsarticle}

\usepackage{mathptmx}
\DeclareMathAlphabet{\mathcal}{OMS}{cmsy}{m}{n}
\usepackage{amsmath}
\usepackage{amssymb}
\usepackage{siunitx}
\usepackage{tabularx}
\newcolumntype{Y}{>{\centering\arraybackslash}X}
   
\usepackage{booktabs}
\usepackage{multirow}
\usepackage{lineno}
\modulolinenumbers[5]
\usepackage[colorlinks=true,linkcolor=blue]{hyperref}
\usepackage[labelfont=bf,font=footnotesize,labelsep=period]{caption}

\journal{journal}

\usepackage{natbib}
\setcitestyle{square}
\biboptions{sort&compress}
\usepackage{titlesec}

\titleformat*{\section}{\Large\bfseries}
\titleformat*{\subsection}{\large\bfseries}
\titleformat*{\subsubsection}{\normalsize\itshape}

\begin{document}

\begin{frontmatter}

\title{ \textbf{Identifying mechanisms driving the early response of triple negative breast cancer patients to neoadjuvant chemotherapy using a mechanistic model integrating \textsl{in vitro} and \textsl{in vivo} imaging data} }

\author[unipv,oden]{Guillermo Lorenzo\corref{correspondingauthor}}
\cortext[correspondingauthor]{
Corresponding author: Guillermo Lorenzo, PhD. Oden Institute for Computational Engineering and Sciences, The University of Texas at Austin, 201 E 24th St, 78712-1229 Austin TX, USA. Email: guillermo.lorenzo@utexas.edu, guillermo.lorenzo@unipv.it
}

\author[oden,livestrong]{Angela M. Jarrett}
\author[vandy1]{Christian T. Meyer}
\author[vandy1,vandy2,vandy3]{Vito Quaranta}
\author[vandy1,vandy2,vandy3]{Darren R. Tyson}
\author[oden,livestrong,ut1,ut2,ut3,mdacc]{Thomas E. Yankeelov}

\address[unipv]{Department of Civil Engineering and Architecture, University of Pavia, Italy}
\address[oden]{Oden Institute for Computational Engineering and Sciences, The University of Texas at Austin, USA}
\address[livestrong]{Livestrong Cancer Institutes, Dell Medical School, The University of Texas at Austin, USA}
\address[vandy1]{Center for Cancer Systems Biology at Vanderbilt, Vanderbilt University, USA}
\address[vandy2]{Department of Biochemistry, Vanderbilt University, USA}
\address[vandy3]{Department of Pharmacology, Vanderbilt University School of Medicine, USA}
\address[ut1]{Department of Biomedical Engineering, The University of Texas at Austin, USA}
\address[ut2]{Department of Diagnostic Medicine, The University of Texas at Austin, USA}
\address[ut3]{Department of Oncology, The University of Texas at Austin, USA}
\address[mdacc]{Department of Imaging Physics, The University of Texas MD Anderson Cancer Center, USA}

\begin{abstract}
Neoadjuvant chemotherapy (NAC) is a standard-of-care treatment for locally advanced triple negative breast cancer (TNBC) before surgery. 
The early assessment of TNBC response to NAC would enable an oncologist to adapt the therapeutic plan of a non-responding patient, thereby improving treatment outcomes while preventing unnecessary toxicities. 
To this end, a promising approach consists of obtaining \textsl{in silico} personalized forecasts of tumor response to NAC \textsl{via} computer simulation of mechanistic models constrained with patient-specific magnetic resonance imaging (MRI) data acquired early during NAC. 
Here, we present a model featuring the essential mechanisms of TNBC growth and response to NAC, including an explicit description of drug pharmacodynamics and pharmacokinetics. 
As longitudinal \textsl{in vivo} MRI data for model calibration is limited, we perform a sensitivity analysis to identify the model mechanisms driving the response to two NAC drug combinations: doxorubicin with cyclophosphamide, and paclitaxel with carboplatin. 
The model parameter space is constructed by combining  patient-specific MRI-based \textsl{in silico} parameter estimates and \textit{in vitro} measurements of pharmacodynamic parameters obtained using time-resolved microscopy assays of several TNBC lines.
The sensitivity analysis is run in two MRI-based scenarios corresponding to a well-perfused and a poorly-perfused tumor.
Out of the 15 parameters considered herein, only the baseline tumor cell net proliferation rate along with the maximum concentrations and effects of doxorubicin, carboplatin, and paclitaxel exhibit a relevant impact on model forecasts (total effect index, $S_T>$0.1). 
These results dramatically limit the number of parameters that require \textsl{in vivo} MRI-constrained calibration, thereby facilitating the clinical application of our model.
\end{abstract}


\end{frontmatter}

\noindent\textbf{Highlights}
\begin{itemize}
\item We present a mechanistic model of TNBC growth and response to NAC
\item The model can be personalized with patient-specific MRI data and drug regimens
\item The model parameter space is constructed by combining \textsl{in silico} and experimental data
\item A sensitivity analysis shows that only a few parameters need patient-specific calibration
\item Computer simulations of the model capture the different clinical NAC outcomes
\end{itemize}
\rule{\linewidth}{0.1mm}


\section{Introduction}\label{intro}

Neoadjuvant chemotherapy (NAC) is widely considered a standard-of-care treatment for stage II-III, locally advanced triple negative breast cancer (TNBC) prior to surgery \cite{Liu2010,Zardavas2015,Waks2019}.
NAC usually consists of one or two consecutive chemotherapeutic regimens delivered over the course of 4 to 6 months. 
Each of these regimens may involve one or two cytotoxic drugs, which induce tumor cell death (e.g., doxorubicin, paclitaxel).
Hence, the use of NAC in TNBC patients aims to reduce the tumor size, which increases the success rate of breast conservation surgery \cite{King2015,Asselain2018,Petruolo2021}. 
NAC may also completely eliminate the tumor, an outcome which is known as a pathological complete response (pCR).
Importantly, patients who achieve pCR in the neoadjuvant setting are known to have a significantly better prognosis and recurrence-free survival, whereas patients who have residual disease after NAC are at increased risk of early recurrence and death \cite{Minckwitz2012,Spring2020,Yau2022}.
Furthermore, NAC also offers an initial opportunity to treat micrometastases  \cite{Liu2010,Buchholz2003,Kuerer1999,Mathiesen2012}.

The early determination of response to NAC would enable the treating oncologist to adapt the therapeutic regimen of a non-responding patient (e.g., by changing the prescribed drugs as well as their dosage and schedule).
This treatment adjustment could improve treatment outcomes while avoiding unnecessary toxicities \cite{Barker2009,Minckwitz2013}, which may entail an increased likelihood of hospitalization, cardiac damage, leukemia, and even death \cite{Zardavas2015,Waks2019}.
In particular, the early optimization of NAC would be especially advantageous in improving the therapeutic success in TNBC, which usually shows an adverse prognosis and is difficult to treat effectively \cite{Waks2019,Gupta2020}.
Additionally, accurately predicting response to NAC would enable identification of exceptional responders who might benefit from treatment de-escalation, including the possibility of non-surgical management of their disease \cite{Veeraraghavan2017,Spring2022,Gupta2022,Heil2020}. 
To definitively establish that switching therapy significantly improves patient outcomes, accurate methods for predicting response early in the course of NAC are required.
However, the existing approaches to assess NAC response do not provide an accurate early prediction of therapeutic outcome.
While imaging-based changes in tumor size metrics (e.g., longest axis, tumor volume) are extensively used in clinical practice \cite{Eisenhauer2009,Hylton2012}, these measurements generally lag behind biological responses to NAC because they do not account for therapeutically-induced changes in important tumor characteristics, such as the quantity and distribution of tumor cells and vasculature \cite{Pickles2006,Tudorica2016}.
Hence, tumor size-based methods cannot definitively establish changes until the patient has received several treatment cycles.
Several tissue-based biomarkers have also been proposed to predict the response of the prescribed NAC on the breast cancer cells of individual patients (e.g., mitotic index, Ki67 index, HER2 expression, Oncotype DX, MammaPrint) \cite{PenaultLlorca2016,Burcombe2006,Nicolini2018,Duffy2017}. 
However, these biomarkers are subtype-specific and require an invasive biopsy to retrieve the tissue samples, which makes them prone to sampling errors due to tumor heterogeneity \cite{Rakha2007,Robertson2019,Qi2021}.

Alternatively, imaging-based computational tumor forecasting has been actively investigated to obtain early predictions of patient-specific pathological and therapeutic outcomes that can guide clinical decision-making for different tumor types \cite{Yankeelov2013,Mang2020,Lorenzo2023,Wang2009,Lipkova2019,Lorenzo2022,Hormuth2021,Chen2012,Wong2017,Liu2014}.
In particular, spatiotemporally-resolved computational forecasts of breast cancer response to NAC have shown promise in predicting the therapeutically-induced reduction of tumor burden at the conclusion of NAC for individual patients \cite{Jarrett2018,Jarrett2020,Wu2022b}.
This forecasting technology leverages biologically-inspired mechanistic models that describe the key mechanisms driving the therapeutic response of the patient's tumor to the prescribed NAC regimen by means of a set of partial differential equations  \cite{Lorenzo2023,Yankeelov2013,Jarrett2021}.
The personalization of these mechanistic models can be achieved by adjusting their parameters for each individual patient.
Towards this end, longitudinal anatomic and quantitative magnetic resonance imaging (MRI) data can provide  patient-specific, spatiotemporally-resolved information on tumor cell density, the local anatomy of the host tissue, and vasculature over the three-dimensional geometry of the affected breast \cite{Lorenzo2023,Jarrett2021,Yankeelov2013}.
Hence, if the model is personalized with MRI data acquired early in the course of NAC, ensuing computer simulations can provide a patient-specific forecast of therapeutic response at the conclusion of the prescribed NAC regimen and prior to surgery, thereby enabling an early prediction of pCR status that could guide the treating oncologist in implementing potential treatment adjustments for individual patients. 
Indeed, patient-specific computational forecasts of breast cancer response to alternative NAC plans have shown that not all patients may benefit from the same NAC regimen and that these computational predictions may provide guidance in defining superior personalized treatment strategies \cite{Jarrett2020,Wu2022}.

While anatomic and quantitative MRI data may provide rich spatially-resolved information about the architecture and physiology of the tumor and host tissue, there are two central limitations in the use of these data types for the patient-specific parameterization of spatiotemporal mechanistic models of tumor growth and therapeutic response: the scarcity of longitudinal MRI measurements in the clinical setting, and the inability to probe the tumor biology at cellular scale \cite{Lorenzo2023,Jarrett2021,Kazerouni2020,Mang2020}.
The limited number of imaging datasets over time poses a central challenge for the patient-specific determination of key model parameters with a dynamic nature (e.g., tumor cell proliferation, mobility, drug-induced cytotoxic effects).
Global variance-based sensitivity analysis \cite{Saltelli2008,Lorenzo2023} provides a rigorous, systematic approach to identify the dominant model parameters that characterize the mechanisms driving tumor growth and response to treatment within a mechanistic model.
These driving mechanisms require patient-specific parameterization, but the rest of the model parameters can be fixed to a constant value for all patients.
Hence, global variance-based sensitivity analysis enables the definition of reduced models that approximate the overall tumor dynamics using the original model formulation, but only requiring the patient-specific identification of a reduced number of dominant parameters.
Furthermore, the cytotoxic action of common chemotherapeutic agents is known to vary at cellular level \cite{Zoli1995,Kim2018}, ultimately depending on the specific genetic profile of the cancer cell subpopulations within a patient's tumor and their evolution during NAC \cite{Dressman2006,Balko2014,PenaultLlorca2016,Nicolini2018,Kim2018}.
This heterogeneity in tumor biology can be systematically examined by leveraging high-throughput, time-resolved, automated microscopy assays.
This technology can evaluate \textsl{in vitro} the response of multiple cancer cell cultures (e.g., from standardized lines, animal xenografts, or patient tissue samples) to numerous therapeutic regimens involving one or more drugs delivered with multiple dosages and schedules \cite{Pepperkok2006,McKenna2017,Meyer2019,Howard2022}.
In this experimental setup, established mechanistic models are also leveraged to quantify the pharmacodynamic effect of each drug regimen on tumor dynamics (e.g., on tumor cell proliferation) \cite{Goutelle2008,Meyer2019,Wooten2021}. 
However, the use of this wealth of \textsl{in vitro} data to parameterize \textsl{in vivo} models of tumor growth and treatment response remains an unresolved challenge due to three main issues: the unique intratumoral cellular heterogeneity of an individual patient's tumor \cite{Almendro2013,Kim2018,Balko2014}, local dynamic variations in the \textsl{in vivo} microenvironment of tumor cells directly affecting therapeutic response (e.g., perfusion, mechanical state) \cite{Jain2014,Stylianopoulos2013,Li2014,Tudorica2016,Kazerouni2022}, and the high number of parameters in the formulation of pharmacodynamic models \cite{Goutelle2008,Meyer2019,Wooten2021}.

Here, we propose a multiscale framework that integrates \textit{in vivo} MRI data characterizing the morphology and function of the tumor and host tissue, \textit{in silico} estimates of tumor dynamics during NAC, and \textit{in vitro} measurements of TNBC cell response to NAC drug combinations to constrain and identify the driving mechanisms of organ-scale models of TNBC response to NAC.
Hence, our framework can determine the model parameters requiring patient-specific calibration by exclusively using \textsl{in vivo} MRI data from individual patients acquired early during the course of NAC to ultimately perform accurate predictions of therapeutic outcomes. 
To this end, we also present a new organ-scale mechanistic model of breast cancer growth and response to NAC that couples the essential mechanisms underlying tumor dynamics with an explicit formulation of the prescribed NAC drugs' pharmacokinetics and pharmacodynamics \cite{Jarrett2018,Jarrett2020,Lorenzo2023}.
Given that NAC regimens usually consist of drug combinations, we adopt a recent pharmacodynamics model \cite{Meyer2019} that generalizes prior formulations of the pharmacodynamics of multidrug regimens and successfully describe the combined effects of multiple drug pairs over a range of tumor cell lines.
The resulting organ-scale model features fifteen parameters that could be eligible for personalized parameterization with \textsl{in vivo} MRI data to perform tumor forecasting.
To reduce this parameter set, we identify the driving parameters of our model using a global variance-based sensitivity analysis \cite{Saltelli2008,Lorenzo2023} over a parameter space constructed by integrating \textsl{in silico} estimates informed by patient-specific \textit{in vivo} MRI data and \textsl{in vitro} pharmacodynamic measurements.
In particular, we experimentally constrain the parameters accounting for drug pharmacodynamics \textit{via} time-resolved, high-throughput, automated microscopy assays that capture drug-induced changes in the proliferation rates of four TNBC cell lines (HCC1143, SUM149, MDAMB231, and MDAMB468).
The resulting \textsl{in vitro} parameter ranges are then scaled to clinically-relevant values through computer simulations with our mechanistic model.
In this work, we focus on two commonly used NAC regimens:  doxorubicin plus cyclophosphamide, and  paclitaxel plus carboplatin \cite{Waks2019,Gupta2020,Poggio2018,Jarrett2020}.
Additionally, the sensitivity analysis is carried out in two representative scenarios of breast cancer in the clinical setting corresponding to a well-perfused and a poorly-perfused tumor, which are respectively extracted from \textit{in vivo} MRI data from two TNBC patients.
Furthermore, we show that computer simulations of the original and reduced model emanating from the sensitivity analysis can equally reproduce the clinical outcomes of NAC under the two NAC drug combination explored in this study, ranging from pCR to progressive disease.

The remainder of this work is organized as follows. 
The next section describes the data used in this study, the mechanistic model of breast cancer growth and NAC response, the numerical methods for the spatiotemporal discretization of the model to enable computer simulations, and the formulation of the global variance-based sensitivity analysis.
Then, we present the results of this sensitivity analysis along with a set of illustrative computer simulations of the model.
Finally, we discuss the results and implications of our work, its limitations, and potential avenues of research in future studies.

\section{Methods}

\subsection{MRI data}\label{mri}

The MRI datasets for this study were obtained from a database of patient-specific, longitudinal anatomic and quantitative MRI \textsl{in vivo} measurements of breast cancer response to NAC at multiple time points before and during the course of the neoadjuvant regimen.
This database has been presented in detail in Refs.~\cite{Jarrett2018,Jarrett2020,Jarrett2021}, so here we present only the salient details and briefly summarize the corresponding preprocessing methods.
For each breast cancer patient, five MRI data types were acquired at each scan session: (1) a pre-contrast T1 map, (2) a pre-contrast B1 field map to correct for radiofrequency inhomogeneity, (3) diffusion-weighted MRI (DW-MRI) data, (4) dynamic contrast-enhanced MRI (DCE-MRI) data, and (5) two high-resolution, T1-weighted anatomical scans (pre- and post-contrast). 
Then, the preprocessing pipeline of these MRI datasets consisted of four steps.
First, an intra-scan rigid registration was applied to the MRI datasets collected within each scan session to correct for motion. 
Second, the tumor regions of interest (ROIs) were identified based on post-contrast scans using a fuzzy c-means algorithm \cite{Wu2020}. 
Additionally, tissue properties related to perfusion-permeability of the vasculature were quantified by analyzing the DCE-MRI data with the Kety-Tofts model, which is a standard model describing the exchange of contrast agent between the plasma and tissue spaces \cite{Yankeelov2007}. 
The DW-MRI data were also analyzed to return maps of the apparent diffusion coefficient (ADC) of water.
Third, an inter-scan registration was leveraged to align the images and calculated maps across all of the imaging sessions into a common geometric domain unique to each patient. 
This inter-scan registration relied on a non-rigid algorithm with a constraint that preserves the tumor volumes at each time point.
Finally, the last step of the preprocessing pipeline consisted of the calculation of the imaging-based quantities to be used within the spatiotemporal mechanistic model (see Section~\ref{model} and Refs.~\cite{Jarrett2018,Jarrett2020,Weis2013,Weis2015,Weis2017,Lorenzo2023}).
In paticular, the tumor cell density maps were approximated from the voxel-based ADC values within the tumor ROI \cite{Atuegwu2011,Atuegwu2013,Jarrett2018,Jarrett2020,Hormuth2021,Lorenzo2023},
fibroglandular and adipose tissues were segmented based on enhancement in the DCE-MRI data by leveraging a k-means clustering algorithm, and a normalized perfusion map representing the spatial distribution of blood volume fraction within the breast tissue was calculated using the area under the dynamic curve (AUC) for each voxel in the DCE-MRI data \cite{Jarrett2020}.

\begin{figure*}[!t]
\centering
\includegraphics[width=\linewidth]{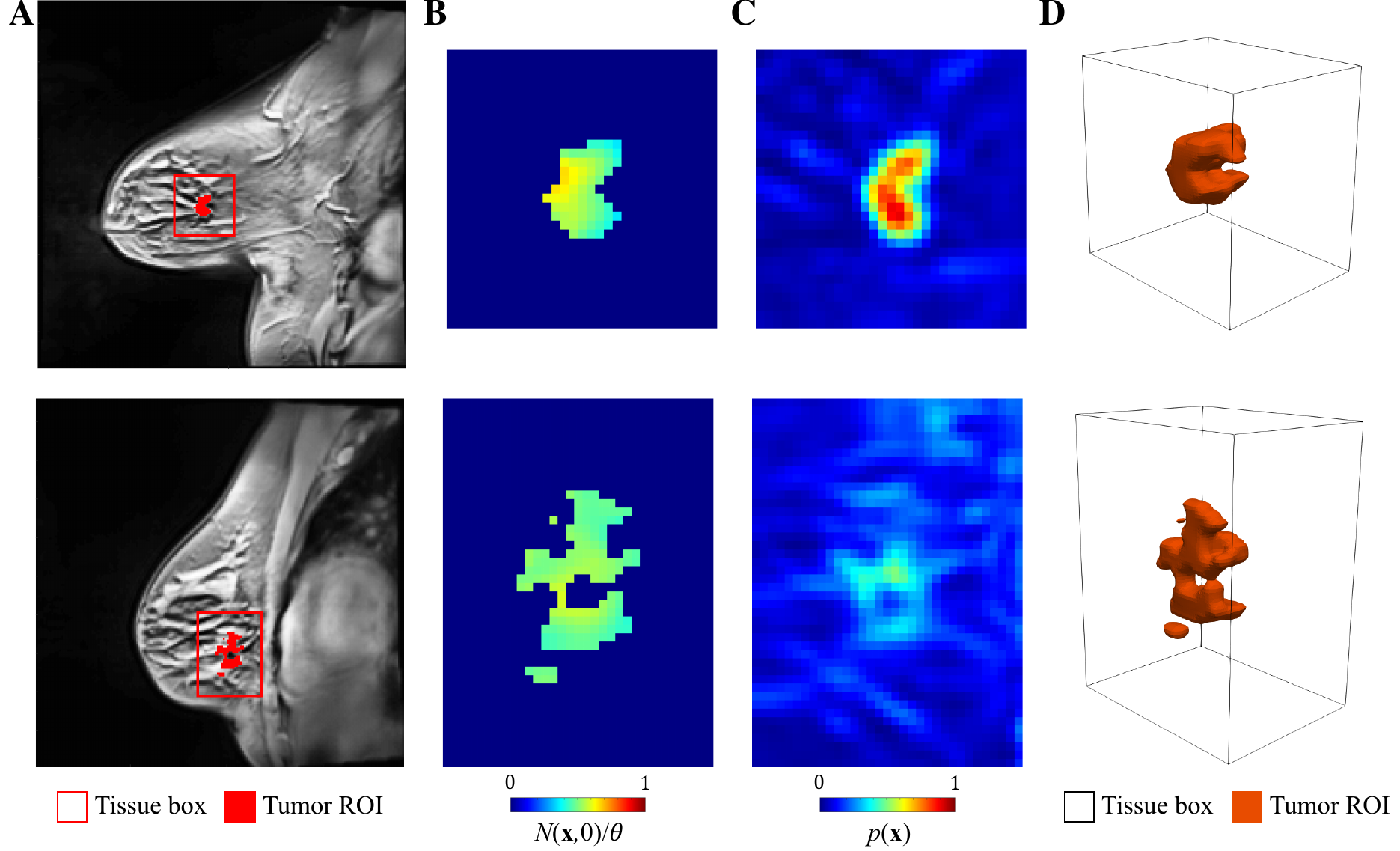}
\caption{\textbf{Patient-specific, MRI-informed tumor scenarios for the sensitivity analysis}. 
Panels A-D show the \textit{in vivo} MRI measurements characterizing the two TNBC cases considered in this study: a well-perfused tumor (top row) and a poorly-perfused tumor (bottom row).
Panel A presents a contrast-enhanced $T_1$-weighted sagittal image showing the patient's breast anatomy and the tissue box containing the tumor region of interest (ROI) for each scenario.
Panel B shows the corresponding tumor cell density maps over a sagittal section of the tissue box.
These tumor cell density maps were calculated from ADC measurements obtained from DW-MRI data and normalized with respect to the tissue carrying capacity ($N(\mathbf{x},t)/\theta$, see Section~\ref{tumdym}).
Panel C depicts the local perfusion map over the same sagittal section of the tissue box for each tumor case. 
These perfusion maps were computed from DCE-MRI data ($p(\mathbf{x})$, see Section~\ref{pkpd}).
Finally, panel D shows a 3D rendering of each tumor ROI within the corresponding tissue box. }
\label{fig_tboxes}
\end{figure*}

The well-perfused and poorly-perfused breast tumor scenarios considered in the sensitivity analysis performed in this work were extracted from the MRI datasets collected for two TNBC patients from the aforementioned database.
While previous computational modeling and forecasting efforts of breast cancer response to NAC consider the tumor dynamics over the patient-specific anatomy of the affected breast \cite{Jarrett2018,Jarrett2020,Jarrett2021,Atuegwu2011,Atuegwu2013,Weis2013,Weis2017}, here we consider a box of breast tissue from each patient including their tumor as well as a surrounding region of healthy tissue.
Figure~\ref{fig_tboxes} shows a 3D rendering and representative sagittal sections of the tissue box of the well-perfused and poorly-perfused breast tumor scenarios, along with their original location in the patient's breast. 
The rationale for selecting this reduced geometry is to minimize the computational resources for the sensitivity analysis, which requires a large number of model simulations (see Section~\ref{sa}).
In particular, we estimated that the box size for each scenario was sufficient to accommodate all potential outcomes of NAC for the varying parameter combinations used within the sensitivity analysis (i.e., ranging from tumor elimination  to consistent growth at the end of treatment), while also limiting the computational resources required to perform a model simulation for each parameter combination. 

\subsection{High-throughput, automated microscopy data}\label{invitro}

Four TNBC cell lines representing different TNBC subtypes \cite{Lehmann2016} were used in this study. 
Each cell line was engineered to express histone 2B-monomeric red fluorescent protein 1 (H2BmRFP1) using recombinant lentiviruses as described previously \cite{Tyson2012}: HCC-1143 (BL1), MDA-MB-231 (M), MDA-MB-468 (BL1), and SUM-149 (BL2). 
Sum-149 cells were cultured in Ham's F-12 medium containing 5\% fetal bovine serum, 1 $\mu$g/ml hydrocortisone and 5 $\mu$g/ml insulin. 
All other cell lines were cultured in Dulbecco's Modified Eagle's Medium (DMEM) containing 10\% fetal bovine serum.
Doxorubicin HCl (Pfizer, NDC 0069-3032-20) was obtained from the Vanderbilt Clinic, while
4-hydroperoxy cyclophosphamide (also known as perfosfamide; the active metabolite form of the prodrug cyclophosphamide), carboplatin, and paclitaxel were purchased from MedChemExpress (cat\# HY-117433, HY-17393, and HY-B0015, respectively). 
Drug combination studies were performed in the Vanderbilt High Throughput Screening Facility as previously described \cite{Meyer2019}. 
Briefly, cells were seeded at approximately 400 cells per well in 384-well plates and allowed to adhere overnight. 
A preliminary image of each plate was taken approximately 8 hours after seeding to verify sufficient cells for each experiment. 
Fluorescence microscopy images were taken on an ImageXpress Micro XL (Molecular Devices). 
Medium containing drugs and 5 nM Sytox Green (to detect dead cells) was added and replaced after 72 hours. 
Cells were imaged over approximately 120 hours of drug exposure. 
Cell counts were determined using custom-written image segmentation software developed in Python using scikit-image \cite{VanderWalt2014} and run in parallel using RabbitMQ/Celery.
Changes in viable cell count over time were used to extract drug-induced proliferation (DIP) rates as previously described \cite{Harris2016,Lubbock2021}. 
Finally, DIP rate values for each drug-drug or single agent drug condition were assessed for any synergistic activity using the multi-dimensional synergy of combinations (MuSyC) formalism \cite{Meyer2019}.
Supplementary Tables S1 and S2 summarize the distribution of experimental measurements of the MuSyC pharmacodynamic parameters in the four TNBC cell lines treated with doxorubicin plus cyclophosphamide and paclitaxel plus carboplatin, respectively. 

\subsection{Mechanistic model of breast cancer dynamics and NAC response}\label{model}

\begin{figure*}[!t]
\centering
\includegraphics[width=\linewidth]{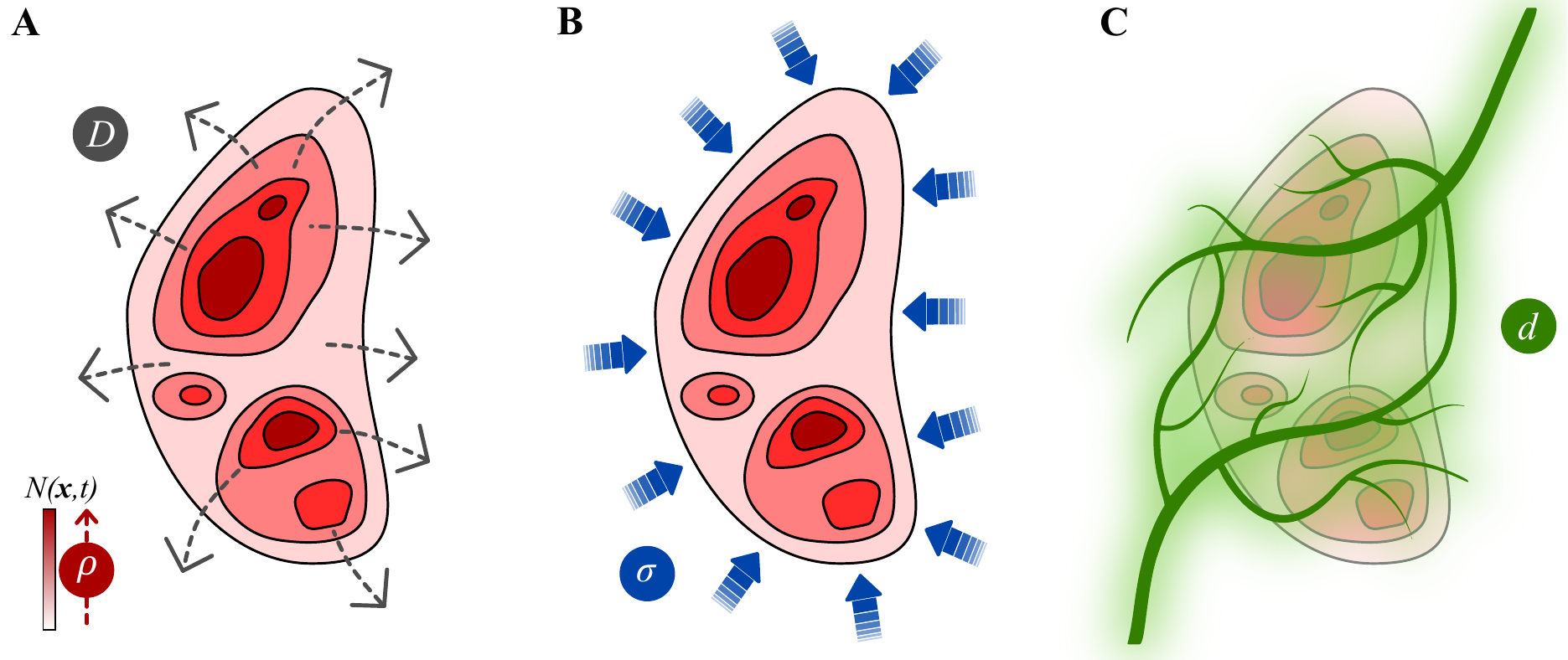}
\caption{\textbf{Mechanistic modeling of breast cancer growth and response to NAC}. 
Panel A illustrates the baseline tumor dynamics in terms of tumor cell density $N(\mathbf{x},t)$ as a combination of mobility and net proliferation, as shown in Eq.~\eqref{rd}.
We model tumor cell mobility with a diffusion operator that is characterized by a tumor cell diffusion coefficient $D$. 
This formulation aims to capture the expansion of the tumor over the surrounding host tissue.
The net proliferation of the tumor cells is formulated with a logistic term, which is driven by the net proliferation rate $\rho$ and locally increases the tumor cell density over time.
Panel B shows that, as solid tumors deform the host tissue in which they are developing, this process induces mechanical stress fields $\sigma$ that are known to inhibit tumor growth.
In our model, we constrain tumor cell mobility with the local mechanical stress field  \textit{via} Eq.~\eqref{dmech}.
Panel C shows that the NAC drugs considered in this study reach the tumor through the bloodstream after being delivered through intravenous injection. 
Since these drugs have a cytotoxic action on tumor dynamics (i.e., they induce tumor cell death), Eq.~\eqref{msyc} locally modifies the tumor cell net proliferation according to the joint effect of the local concentration $d$ of each drug pair during the course of NAC.
In particular, our model couples the pharmacokinetics and pharmacodynamics of the drugs in each NAC regimen \textit{via} Eqs.~\eqref{msyc}-\eqref{pk}, also accounting for drug synergy interactions according to the MuSyC framework (see Fig.~\ref{fig_pkpd}).
Panel C was created using BioRender.}
\label{fig_model}
\end{figure*}

Our mechanistic modeling framework relies on a biologically-inspired approach that aims at describing the main spatiotemporal mechanisms underlying tumor response to NAC at organ scale, such that the driving phenomena in the model can be parameterized for each individual patient by leveraging longitudinal \textsl{in vivo} MRI measurements collected before and during the course of NAC \cite{Jarrett2018,Jarrett2020,Jarrett2021,Yankeelov2013,Lorenzo2023}. 
In particular, we couple a previous mechanically-constrained mechanistic model of breast cancer response to NAC \cite{Weis2013,Weis2015,Weis2017} with a pharmacodynamics model that explicitly captures the combined effect on tumor cell proliferation of drug pairs in NAC regimens \cite{Meyer2019,Wooten2021}, while also accounting for drug pharmacokinetics \cite{Jarrett2018,Jarrett2021}.
Figure~\ref{fig_model} illustrates the mechanisms in the model, which we now describe in detail.

\subsubsection{Tumor dynamics}\label{tumdym}

Let $\mathbf{x}$ denote the position vector within the spatial domain $\Omega$, which consists of a box of breast tissue including the tumor (see Section~\ref{mri}), and let $t$ denote time within the interval $[0,T]$, such that $T$ represents the time horizon when model outcomes will be assessed (e.g., after the completion of NAC).
We model breast cancer dynamics in terms of tumor cell density $N(\mathbf{x},t)$ as

\begin{equation}\label{rd}
\frac{\partial N}{\partial t} = \nabla\cdot\left( D\nabla N \right)  + \rho N \left( 1 - \frac{N}{\theta} \right),
\end{equation}

\noindent The right-hand-side of Eq.~\eqref{rd} describes the dynamics of  $N(\mathbf{x},t)$ as a combination of tumor cell mobility and net proliferation by leveraging a diffusion operator and a logistic term, respectively.
In this equation, $D(\mathbf{x},t)$ denotes the tumor cell diffusivity, which quantifies tumor cell mobility.
Additionally, $\rho(\mathbf{x},t)$ represents the tumor cell net proliferation rate, which encompasses the baseline balance of tumor cell proliferation and death as well as the therapeutically-induced tumor cell death due to the cytotoxic action of the prescribed drugs.
Hence, a positive value of $\rho(\mathbf{x},t)$ contributes to tumor growth, whereas a negative value leads towards tumor elimination.
Finally, $\theta$ denotes the breast tissue carrying capacity, which represents the maximum admissible tumor cell density at any point $\mathbf{x}\in\Omega$.
The reaction-diffusion modeling paradigm in Eq.~\eqref{rd} has been successfully employed to represent and forecast the dynamics of breast cancer growth and response to NAC of individual patients \cite{Jarrett2018,Jarrett2020,Weis2013,Weis2015,Roque2018}, as well as the development and treatment response of other tumors \cite{Hormuth2021,Lipkova2019,Wang2009,Neal2013,Scheufele2021,Chen2012,Wong2017,Liu2014,Lorenzo2022}.
Following this modeling paradigm, Eq.~\eqref{rd} is coupled with the zero-flux boundary condition $\nabla N \cdot \mathbf{n} = 0$ on $\Gamma=\partial\Omega$, where $\mathbf{n}$ is an outward unit vector normal to the boundary of the tissue box  (see Section~\ref{mri}). 
Hence, this boundary condition is consistent with the assumption that the tissue box suffices to spatially hold the diverse spectrum of the dynamics of breast cancer response to be investigated during the sensitivity analysis in each tumor scenario.

\subsubsection{Mechanical coupling}\label{mechcoup}

As the tumor grows, the increasing tumor cell density induces the deformation and accumulation of mechanical stress in the host tissue \cite{Helmlinger1997,Jain2014,Fraldi2018,Stylianopoulos2013}.
This phenomenon is usually termed the tumor mass effect.
Since mechanical stress is known to exert an inhibitory effect on tumor growth \cite{Helmlinger1997,Jain2014,Stylianopoulos2013,Fraldi2018}, we constrain the tumor cell diffusion coefficient  $D(\mathbf{x},t)$ in Eq.~\eqref{rd} with the local tumor-induced mechanical stress field in the breast:

\begin{equation}\label{dmech}
D(\mathbf{x},t) = D_0 e^{-\gamma_N\sigma_v(\mathbf{x},t)}.
\end{equation}

\noindent In Eq.~\eqref{dmech}, $D_0$ represents the tumor cell diffusivity in the absence of mechanical inhibition, $\gamma_N$ denotes an empirical coupling constant, and $\sigma_v(\mathbf{x},t)$ is the von Mises stress, which is a scalar stress metric calculated as

\begin{equation}\label{vms}
\sigma_v = \left( \sigma^2_{11}+\sigma^2_{22}+\sigma^2_{33} - \sigma_{11}\sigma_{22} - \sigma_{22}\sigma_{33} -\sigma_{11}\sigma_{33} + 3 \left( \sigma^2_{12} + \sigma^2_{23} + \sigma^2_{13} \right)  \right)^{1/2},
\end{equation}

\noindent where $\sigma_{ij}$ (with $i,j=1,2,3$) are the components of the second-order mechanical stress tensor $\boldsymbol{\sigma}(\mathbf{x},t)$.
This mechanically-coupled approach has been shown to render superior predictions of breast tumor dynamics during NAC with respect to a baseline model without mechanics \cite{Weis2013,Weis2015}, and it has also been adopted in patient-specific, mechanically-constrained models of brain and prostate cancer \cite{Hormuth2021,Lorenzo2019,Lorenzo2020}.

To calculate the mechanical stress tensor $\boldsymbol{\sigma}(\mathbf{x},t)$, we assume that the tumor-induced mechanical deformation of breast tissue is sufficiently slow to disregard inertial effects and that, hence, it is governed by quasistatic equilibrium  \cite{Weis2013,Weis2015,Weis2017,Hormuth2021,Lorenzo2019,Lorenzo2020,Mang2020,Liu2014}, which is modeled as

\begin{equation}\label{mecheq}
\mathbf{\nabla}\cdot\boldsymbol{\sigma} = \mathbf{0}.
\end{equation}

\noindent 
Additionally, we set Winkler-inspired boundary conditions $\mathbf{\sigma n} = -k_w\mathbf{u}$ in $\Gamma=\partial\Omega$, where $\mathbf{u}$ is the displacement vector and $\mathbf{n}$ is an outward unit vector normal to the tissue box boundary (see Section~\ref{mri}).
These boundary conditions aim at modeling the mechanical constraint to tumor growth imposed by the tissues surrounding the box domain in each tumor scenario \cite{Lorenzo2019,Lorenzo2020}.
The parameter $k_w$ in Eq.~\eqref{mecheq} quantifies the mechanical stresses induced by the neighboring tissues per unit of the boundary displacement.

Linear elasticity has been widely used to describe the tumor-induced deformation of the host tissue in clinically-oriented models of tumor growth and treatment response  \cite{Weis2013,Weis2015,Weis2017,Hormuth2021,Lorenzo2019,Lorenzo2020,Mang2020,Liu2014}.
Additionally, the breast is a histologically heterogeneous organ composed of fibroglandular and adipose tissue, with the former type of tissue exhibiting a higher elastic modulus \cite{Weis2013,Weis2015,Weis2017}.
Hence, breast tissue can be modeled as a linear elastic, heterogeneous, isotropic material with a constitutive equation given by

\begin{equation}\label{sigma}
\boldsymbol{\sigma} = \lambda\left( \mathbf{\nabla}\cdot\mathbf{u} \right)\mathbf{I} + \mu  \left( \mathbf{\nabla u} + \mathbf{\nabla u}^T \right) - g_N  \frac{N}{\theta} \mathbf{I},
\end{equation}

\noindent where $\boldsymbol{\sigma}(\mathbf{x},t)$ is the mechanical stress tensor, $\mathbf{u}(\mathbf{x},t)$ is the displacement vector, $\mathbf{\nabla}^s\mathbf{u}$ is the symmetric gradient of the displacement field (i.e., the strain tensor in linear elasticity, $\boldsymbol{\varepsilon}=\mathbf{\nabla}^s\mathbf{u}=\left( \mathbf{\nabla u} + \mathbf{\nabla u}^T \right)/2$), $\mathbf{I}$ is the second-order identity tensor, and $\lambda(\mathbf{x})$ and $\mu(\mathbf{x})$ are the local Lam\'{e} coefficients.
These mechanical parameters are related to the local tissue Young elastic modulus $E$ and Poisson's ratio $\nu$ as $\lambda = {E\nu}/{((1+\nu)(1-2\nu))}$ and $\mu = {E}/{(2(1+\nu))}$.
Thus, in Eq.~\eqref{sigma}, the first two terms in the right-hand side correspond to the usual linear elasticity constitutive equation while the third term describes the tumor mass effect as a growth-induced phenomenon, which is characterized by the tumor-induced solid stress $g_N$ and is proportional to the normalized tumor cell density $N(\mathbf{x},t)/\theta$  \cite{Weis2013,Weis2015,Weis2017,Lorenzo2019,Lorenzo2020,Hormuth2021,Mang2020,Liu2014}.

\subsubsection{Pharmacodynamics and pharmacokinetics of NAC drugs}\label{pkpd}

Standard NAC drug combinations induce tumor cell death, which ultimately reduces or even eliminates the patient's tumor burden \cite{Zardavas2015,Waks2019,Minckwitz2012,Spring2020,Yau2022}.
The effect of NAC drugs on tumor cells is usually quantified \textsl{in vitro} with concentration-dependent pharmacodynamic models, which can further accommodate synergistic effects for multidrug regimens (see Section~\ref{invitro}) \cite{Goutelle2008,Meyer2019,Wooten2021}.
However, the relatively large number of parameters in these pharmacodynamic models usually precludes their use in organ-scale clinically-oriented models because the patient-specific clinical and imaging data available for their determination is insufficient.
The standard approach to account for the effect of chemotherapeutic drugs in mechanistic models of tumor growth and treatment response in the clinical setting consists of introducing a term to counteract the proliferation mechanism in the baseline model.
For example, some models extend the equation governing tumor dynamics (e.g., Eq.~\eqref{rd}) with a reaction term proportional to the tumor burden and the drug concentration, while others opt to directly adjust the net proliferation rate with a similar concentration-dependent term after the delivery of each drug dose \cite{Jarrett2018,Jarrett2020,Colli2020,Howard2022,Kohandel2007,McKenna2017,Bogdanska2017,Hinow2009,West2019}.

We propose to explicitly couple a full pharmacodynamic model with the net proliferation rate $\rho(\mathbf{x},t)$ in Eq.~\eqref{rd}.
Our underlying assumptions are: (1) the same pharmacodynamic paradigm can be used to describe the response of TNBC cells \textsl{in vitro} and \textsl{in vivo}, and (2)  the pharmacodynamic parameters of each NAC drug pair are either in the same range for both \textsl{in vitro} and \textsl{in vivo} scenarios or can be scaled accordingly through computer simulations of our mechanistic model (see Section~\ref{params}).
It is important to note that this last assumption does not mean that we determine the pharmacodynamic parameters for a given NAC drug combination \textsl{in vitro} and then directly use these values within the definition of $\rho(\mathbf{x},t)$ in Eq.~\eqref{rd} to predict the response of a patient's tumor to that NAC regimen.
Rather, we define the value ranges for the pharmacodynamic parameters using the \textsl{in vitro} data in Section~\ref{invitro} and, then, use those ranges to construct the parameter space to constrain our mechanistic model for sensitivity analysis in the clinical setting. 
In Section~\ref{discussion} we argue that these ranges can also be used for personalized calibration using longitudinal MRI data acquired for each individual patient during the course of NAC.

We adjust the proliferation rate $\rho(\mathbf{x},t)$ in Eq.~\eqref{rd} by using a normalized version of the MuSyC model that we leveraged to analyze the response of TNBC cell lines to NAC regimens in Section~\ref{invitro}.
In this approach, we normalize the drug concentrations $d_i(\mathbf{x},t)$ and maximal effects on tumor cell proliferation $E_i$ for each of the two NAC drugs ($i=1,2$) used in the regimens considered in this work (i.e., doxorubicin plus cyclophosphamide, and paclitaxel plus carboplatin).
This normalization serves two purposes to bridge the use of the MuSyC model from \textsl{in vitro} to \textsl{in vivo} scenarios.
First, while drug concentrations can be precisely controlled in experimental settings, the corresponding values within the tumor and host tissue of an individual patient cannot be measured with routine clinical imaging methods.
However, our mechanistic model does not necessarily require the precise value of drug concentrations, but rather their relative temporal change following the delivery of each dose (e.g., due to tumor consumption and elimination)   \cite{Jarrett2018,Jarrett2020,Colli2020,Howard2022,Kohandel2007,McKenna2017,Bogdanska2017,Hinow2009,West2019}.
Second, the maximal drug effects may differ between an \textsl{in vitro} setting and a patient-specific \textsl{in vivo} scenario.
For example, standard NAC drugs for TNBC have been observed to reduce the net proliferation rate of breast tumor cell populations and limit their growth (e.g., see Supplementary Tables S1 and S2 and Refs. \cite{Howard2022,McKenna2017,Zoli1995,Zoli1999}). 
However, the delivery of the same drugs in routine NAC regimens is known to achieve the complete tumor elimination in some breast cancer patients (i.e., a pCR)  \cite{Zardavas2015,Waks2019,Minckwitz2012,Spring2020,Yau2022}.
As for drug concentrations, from a mechanistic modeling perspective it suffices to quantify the drug-induced relative change in tumor cell proliferation to capture the chemotherapeutic effect of a given drug regimen on tumor dynamics in both preclinical and clinical scenarios \cite{Jarrett2018,Jarrett2020,Colli2020,Howard2022,Kohandel2007,McKenna2017,Bogdanska2017,Hinow2009,West2019}.

To construct the normalized MuSyC model, we define $\rho_0$ as the global baseline value of the tumor cell net proliferation before the onset of NAC, and $C_i$ ($i=1,2$) as the half-maximal effective concentration of each NAC drug in a particular regimen (i.e., the drug concentration producing half of the maximal effect for each individual drug).
Then, we define the normalized drug concentrations and maximal effects on tumor cell net proliferation as $\hat{d}_i(\mathbf{x},t)=d_i(\mathbf{x},t)/C_i$ and $\hat{E}_i=E_i/\rho_0$, respectively ($i=1,2$).  
Using the previous definitions, we express the tumor cell net proliferation rate $\rho(\mathbf{x},t)$ from Eq.~\eqref{rd} in terms of the normalized drug concentrations $\hat{d}_1(\mathbf{x},t)$ and $\hat{d}_2(\mathbf{x},t)$ in each NAC drug regimen as

\begin{equation}\label{msyc}
\rho = \rho_0 \frac{ g_0\left(\hat{d}_1,\hat{d}_2\right) + g_1\left(\hat{d}_1,\hat{d}_2\right)\hat{d}_1^{h_1}\hat{E}_1 + g_2\left(\hat{d}_1,\hat{d}_2\right)\hat{d}_2^{h_2}\hat{E}_2 + g_3\left(\hat{d}_1,\hat{d}_2\right)\hat{d}_1^{h_1}\hat{d}_2^{h_2}\hat{E}_3}{ g_0\left(\hat{d}_1,\hat{d}_2\right) + g_1\left(\hat{d}_1,\hat{d}_2\right)\hat{d}_1^{h_1} + g_2\left(\hat{d}_1,\hat{d}_2\right)\hat{d}_2^{h_2} + g_3\left(\hat{d}_1,\hat{d}_2\right)\hat{d}_1^{h_1}\hat{d}_2^{h_2} },
\end{equation}

\noindent where

\begin{equation}\label{g0}
g_0\left(\hat{d}_1,\hat{d}_2\right) = C_1^{h_1} + C_2^{h_2} + \left(C_1\alpha_2\hat{d}_1\right)^{h_1} + \left(C_2\alpha_1\hat{d}_2\right)^{h_2}
\end{equation}

\begin{equation}\label{g1}
g_1\left(\hat{d}_1,\hat{d}_2\right) = C_1^{h_1} + C_2^{h_2} + \left(C_1\alpha_2\hat{d}_1\right)^{h_1} + \alpha_2^{h_1}\left(C_2\hat{d}_2\right)^{h_2}
\end{equation}

\begin{equation}\label{g2}
g_2\left(\hat{d}_1,\hat{d}_2\right) = C_1^{h_1} + C_2^{h_2} + \alpha_1^{h_2}\left(C_1\hat{d}_1\right)^{h_1} + \left(C_2\alpha_1\hat{d}_2\right)^{h_2}
\end{equation}

\begin{equation}\label{g3}
g_3\left(\hat{d}_1,\hat{d}_2\right) = \left(C_1\alpha_2\right)^{h_1} + \left(C_2\alpha_1\right)^{h_2} + \alpha_1^{h_2}\left(C_1\alpha_2\hat{d}_1\right)^{h_1} + \alpha_2^{h_1}\left(C_2\alpha_1\hat{d}_2\right)^{h_2}
\end{equation}

\begin{equation}\label{e3}
\hat{E}_3 = \left(1+\beta\right)\min\left(\hat{E}_1,\hat{E}_2\right) - \beta
\end{equation}

\begin{figure*}[!t]
\centering
\includegraphics[width=\linewidth]{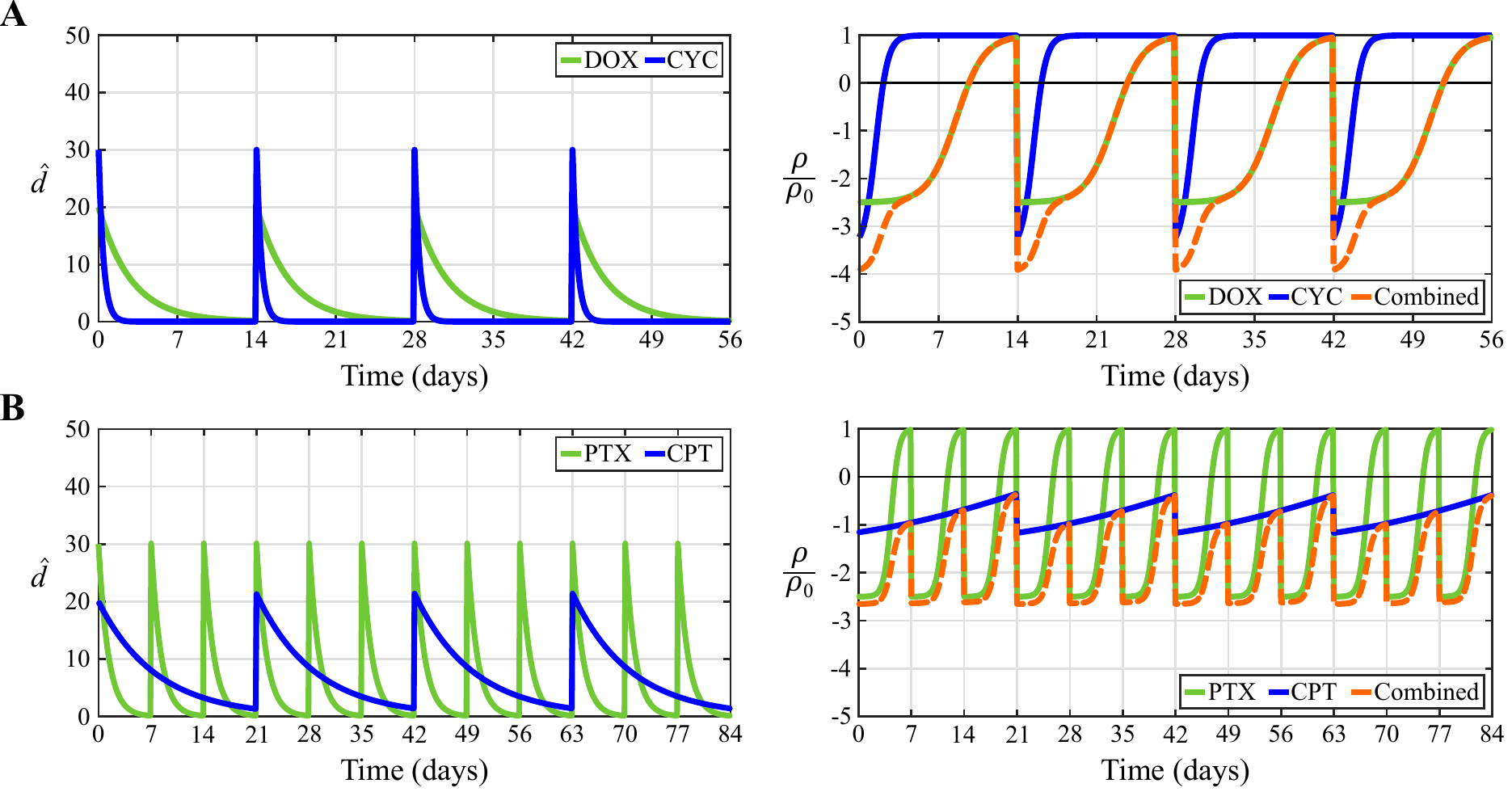}
\caption{\textbf{Pharmacokinetics and pharmacodynamics of the NAC drug pairs}. 
We consider two NAC regimens: four two-week cycles of doxorubicin plus cyclophosphamide (DOX and CYC, panel A), and four three-week cycles of weekly paclitaxel plus three-weekly carboplatin (PTX and CPT, panel B).
In each panel, the left plot illustrates the pharmacokinetics of either drug in the regimen at an arbitrary spatial point of perfused tissue $\mathbf{x}\in\Omega$ (i.e., the temporal change in local drug concentration according to Eq.~\eqref{pk} in Section~\ref{pkpd}).
The peaks in the normalized drug concentration $\hat{d}$ for each drug correspond to the days of delivery.
The right plot in each panel shows the corresponding drug-induced changes in tumor cell proliferation caused by each drug alone as well as in combination. 
In the latter case, synergistic effects emerge according to the MuSyC framework \cite{Meyer2019}.
The drug-induced changes are calculated by combining the pharmacodynamics and pharmacokinetics formulations within our proposed model of breast cancer response to NAC (i.e, Eqs.~\eqref{msyc} and \eqref{pk} in Section~\ref{pkpd}).}
\label{fig_pkpd}
\end{figure*}

\noindent In Eqs.~\eqref{msyc}-\eqref{e3}, $h_i$ and  $\alpha_i$ ($i=1,2$) respectively denote the Hill coefficient and the synergy of potency of each drug in the regimen, which measures the increase in potency for a given concentration when the drugs are delivered in combination.
The parameter $\beta$ is the synergy of efficacy, which defines the additional effect $E_3$ on tumor cell proliferation achieved when the two drugs in the NAC regimen are given concomitantly, as indicated in  Eq.~\eqref{e3}. 
Thus, in our modeling framework, the parameter set comprised of $h_1$, $h_2$, $\hat{E}_1$, $\hat{E}_2$, $C_1$, $C_2$, $\alpha_1$, $\alpha_2$, and $\beta$ characterizes the pharmacodynamics of an NAC drug combination.
Figure~\ref{fig_pkpd} provides an example to illustrate the pharmacodynamics of the two NAC regimens considered in this study.

Following the delivery of each drug dose, pharmacokinetic models describe the temporal dynamics of the drug concentration in the patient's body (e.g., drug infusion, systemic distribution, consumption, and elimination) \cite{Luepfert2005,Buxton2018,Joerger2007,Jonge2005,Joerger2007a,Vijgh1991}.
These pharmacokinetic models result in a terminal decaying phase that dominates after a few hours or days of drug delivery \cite{Speth1988,Kontny2013,Jonge2005,Anderson1996,Ohtsu1995,Gianni1997,Vijgh1991}.
Thus, an exponentially decaying formulation provides a reasonable approximation of the relevant drug pharmacokinetics during the time scale of  chemotherapeutic regimens, since each drug dose is usually delivered every 1 to 3 weeks.
Indeed, this modeling strategy is a common approach to approximate drug pharmacokinetics in mechanistic models of tumor response to systemic therapies   \cite{Jarrett2018,Jarrett2020,Colli2020,Kohandel2007,Bogdanska2017,Wu2022}.
Here, we adopt the exponential decay formulation for drug pharmacokinetics and we further extend it to accommodate the drug distribution across the tumor and breast tissue, as proposed in \cite{Jarrett2018,Jarrett2020}.
Hence, we define the normalized drug concentrations $\hat{d}_i (\mathbf{x},t)$ in Eqs.~\eqref{msyc}-\eqref{g3} as

\begin{equation}\label{pk}
\hat{d}_i (\mathbf{x},t) = \sum_{j=1}^{n_i} \hat{d}_{m,i} p(\mathbf{x})e^{-\gamma_i\left(t-t_{ij}\right)}H(t-t_{ij})
\end{equation}

\noindent where subindices $i$ and $j$ denote each drug in the NAC regimen ($i=1,2$) and the number of each dose in the treatment plan of the $i$-th drug ($j=1,\ldots,n_i$), respectively.
Hence, $t_{ij}$ represents the delivery time of the $j$-th dose of the $i$-th drug, which only contributes to the sum in Eq.~\eqref{pk} for times $t>t_{ij}$ due to the Heaviside function $H(t-t_{ij})$.
The parameter $\hat{d}_{m,i}$ ($i=1,2$) is the normalized maximum drug concentration.
The spatial map $p(\mathbf{x})$ describes blood perfusion in the tumor and the breast tissues, and it is calculated from patient-specific DCE-MRI data (see Section~\ref{mri}).
Additionally, the parameter $\gamma_i$ ($i=1,2$) is the decay rate of each drug in the NAC regimen. 
Figure~\ref{fig_pkpd} depicts an example of the pharmacokinetics of the drugs participating in the two NAC regimens considered in this study.

\subsection{Numerical methods}\label{comp}

\subsubsection{Spatial discretization}\label{discspace}

We discretize in space using Isogeometric Analysis (IGA), which is a recent generalization of the classic Finite Element Method  \cite{Cottrell2009,Lorenzo2023}. 
In particular, our spatial discretization relies on a standard isogeometric Bubnov-Galerkin approach using a three-dimensional $C^1$ quadratic B-spline space \cite{Cottrell2009,Colli2020,Lorenzo2020,Lorenzo2023}.
The isogeometric meshes for the well-perfused and poorly-perfused tumor scenarios were constructed by upsampling the original voxel grid of the corresponding tissue boxes (see Fig.~\ref{fig_tboxes} and Section~\ref{mri}) for numerical accuracy (see Supplementary Methods S1  for further detail).

\subsubsection{Time discretization}\label{disctime}

We partition the time domain $[0,T]$ using a constant time step of $0.25$ days \cite{Weis2017,Jarrett2018}.
The initial time $t=0$ corresponds to the onset of NAC.
We consider a standard doxorubicin plus cyclophosphamide regimen consisting of four two-week cycles in which both drugs are delivered the first day of each cycle \cite{Waks2019,Poggio2018,Jarrett2018,Jarrett2020} (see Fig.~\ref{fig_pkpd}).
Additionally, we consider a paclitaxel plus carboplatin regimen consisting of four three-week cycles in which paclitaxel is administered weekly, while carboplatin is delivered on the first day of each cycle \cite{Poggio2018,Jarrett2018,Jarrett2020} (see Fig.~\ref{fig_pkpd}).
For both regimens, we set the time horizon $t=T$ at the end of the last cycle, such that $T=56$ days for the doxorubicin plus cyclophosphamide regimen and $T=84$ days for the paclitaxel plus carboplatin regimen.

\subsubsection{Numerical solvers}\label{algo}

Given that the tumor dynamics in Eq.~\eqref{rd} corresponds to a transient problem while the mechanical equilibrium in Eq.~\eqref{mecheq} is quasistatic, we adopt a staggered approach to solve the spatiotemporal discretization of the model equations \cite{Lorenzo2019,Lorenzo2020}. 
Since the tumor-driven deformation of the breast tissue  during NAC is a slow process and we are solving our model at four time points during every natural day, we solve the quasistatic mechanical equilibrium every four time steps to reduce the computational time required in the simulations of this study \cite{Lorenzo2019,Lorenzo2020}.
Hence, every four time steps starting at $t=0$, we first solve mechanical equilibrium to update the displacements and then we update the tumor cell density by solving tumor dynamics, while in the remainder time steps we only solve the latter.
We solve tumor dynamics by applying the generalized-$\alpha$ method to the spatiotemporal discretization of Eq.~\eqref{rd}, which leads to a nonlinear system of equations in each time step \cite{Chung1993,Jansen2000,Cottrell2009,Colli2020}.
We linearize this system using the Newton-Raphson method, and we solve the resulting linear systems by means of the generalized minimal residual (GMRES) method with a diagonal preconditioner \cite{Saad1986,Cottrell2009,Colli2020}.
The quasistatic mechanical equilibrium is also solved by directly employing the GMRES algorithm with a diagonal preconditioner on the spatiotemporal discretization of Eq.~\eqref{mecheq}.
The integrals involved in the application of the aforementioned solvers to the spatiotemporal discretization of the model are calculated with standard Gaussian quadrature  \cite{Cottrell2009,Colli2020,Lorenzo2020}.
The numerical methods described in this section and the spatiotemporal discretization of our model were implemented using in-house FORTRAN codes, which were built using the algorithms and recommendations provided in Ref.~\cite{Cottrell2009}.

\subsection{Sensitivity analysis}\label{sa}

\begin{table}[!t]
\centering
\caption{List of parameters considered in the sensitivity analysis and corresponding value ranges. For each drug pair in the NAC regimens, parameters with subindex 1 correspond to the first drug (i.e., doxorubicin and paclitaxel) and parameters with subindex 2 correspond the second drug (i.e., cyclophosphamide and carboplatin). For the parameters that were sampled logarithmically, the reported range corresponds to the admissible values of the decimal logarithm of the parameter. DOX: doxorubicin, CYC: cyclophosphamide, PTX: paclitaxel, CPT: carboplatin.}
\begin{footnotesize}
\setlength\tabcolsep{10pt}
\begin{tabularx}{\linewidth}{@{}Xccccc@{}} 
\toprule[0.5mm]
\multirow{2}{*}{Parameter}  & \multirow{2}{*}{Notation} & \multirow{2}{*}{Units} & \multirow{2}{*}{Sampling} &  \multicolumn{2}{c}{NAC regimen}\\ \cmidrule{5-6}
 & & & & DOX \& CYC & PTX \& CPT \\
\midrule
Baseline tumor cell diffusivity & $D_0$ & mm$^2$/day & Logarithmic & [-6, -1] & [-6, -1]\\
Baseline tumor cell net proliferation & $\rho_0$ & 1/day & Logarithmic & [-3, -1] & [-3, -1]\\
\multirow{2}{*}{Drug Hill coefficients} & $h_1$ & - & Logarithmic & [0, 1.4] & [0, 0.8] \\
                                   & $h_2$ & - & Logarithmic & [-0.4, 0.5] & [-0.5, 0.5] \\
\multirow{2}{*}{Half-maximal effective drug concentrations} & $C_1$ & $\mu$M & Logarithmic & [-3.2, -1.8] & [-3, -2.4] \\
                                   & $C_2$ & $\mu$M & Logarithmic & [-0.9, 3.2] & [-1.2, 1.6] \\
\multirow{2}{*}{Normalized maximal drug effects} & $\hat{E}_{1}$ & - & Regular & [-5, 0.5] & [-3, 0.5] \\
                                   & $\hat{E}_{2}$ & - & Regular & [-5, 0.5] & [-3, 0.5] \\ 
\multirow{2}{*}{Coefficients of drug synergy of potency} & $\alpha_{1}$ & - & Logarithmic & [-4, 3.1] & [-4, 4] \\
                                   & $\alpha_{2}$ & - & Logarithmic & [-4, 1.6] & [-4, 0.4] \\ 
Coefficient of drug synergy of efficacy & $\beta$ & - & Regular & [-0.2, 0.3] & [-0.1, 0.2]\\                                   
\multirow{2}{*}{Normalized maximal drug concentrations} & $\hat{d}_{m,1}$ & - & Logarithmic & [0.7, 3] & [0.7, 3] \\
                                   & $\hat{d}_{m,2}$ & - & Logarithmic & [0.7, 3] & [0.7, 3] \\ 
 \multirow{2}{*}{Drug decay rates} & $\gamma_1$ & 1/day & Regular & [0.3, 0.6] & [0.3, 1.1] \\
                                   & $\gamma_2$ & 1/day & Regular & [1.7, 5.4] & [0.1, 0.2] \\
\bottomrule[0.5mm]
\end{tabularx}
\end{footnotesize}
\label{parspace}
\end{table}

\subsubsection{Methodology}\label{sameth}

We perform a global variance-based sensitivity analysis of our model of breast cancer response to NAC using the approach proposed by Saltelli \textit{et al.} \cite{Saltelli2008,Saltelli2010} to calculate the total-effects index $S_{T,i}$ for each parameter $p_i$ ($i=1,...,n_p$).
This sensitivity metric quantifies the total contribution of variations in each model parameter to the variance of the model outcomes, including both first-order effects and higher-order effects due to interactions with other parameters.
We consider a set of $n_p=15$ model parameters for sensitivity analysis, which are listed in Table~\ref{parspace}.
To calculate the total-effects indices $S_{T,i}$, we first generate two independent random samples of the parameter set  leveraging Latin Hypercube Sampling as provided by \textit{lhsdesign} in MATLAB.
Each of these samples features $n_s=1000$ parameter combinations, which are stored in two $n_s \times n_p$ matrices $\mathbf{A}$ and $\mathbf{B}$, respectively.
Hence, each row of these matrices represents a parameter combination and each column corresponds to a parameter $p_i$ ($i=1,\ldots,n_p$).
We further generate a total of $n_p$ matrices $\mathbf{A}_\mathbf{B}^{(i)}$ ($i=1,\ldots,n_p$) by replacing the $i$-th column from matrix $\mathbf{A}$ with the $i$-th column from matrix $\mathbf{B}$.
Then, we run a total of $n_s(n_p+2)$ simulations of our model for each parameter combination in matrices $\mathbf{A}$, $\mathbf{B}$, and $\mathbf{A}_\mathbf{B}^{(i)}$, $i=1,\ldots,n_p$.

In each time step of the model simulations, we use the spatial map of tumor cell density $N(\mathbf{x},t)$ to compute two model outcomes of interest: the tumor volume $V_T$ and the total tumor cell count $N_T$, which are calculated as

\begin{equation}\label{vt}
V_T=\int_\Omega H\left(N(\mathbf{x},t)-N_{th}\right)\; d\Omega
\end{equation}
\noindent and
\begin{equation}
N_T=\int_\Omega N(\mathbf{x},t) \; d\Omega.
\end{equation}

\noindent These two quantities of interest are used ubiquitously in imaging-based computational models of solid tumor growth and treatment response  \cite{Jarrett2018,Jarrett2020,Jarrett2021,Weis2013,Weis2015,Weis2017,Hormuth2021,Lorenzo2023}.
While $V_T$ is calculated from the spatial region occupied by the tumor, $N_T$ accounts for the heterogeneous spatial distribution of tumor cell density.
In Eq.~\eqref{vt}, $H$ denotes the Heaviside step function and $N_{th}$ is a threshold for the model-calculated tumor cell density to define the tumor segmentation on the simulation results, and ultimately compare it to the corresponding segmentation from the patient's MRI data  \cite{Jarrett2018,Jarrett2020,Jarrett2021,Weis2013,Weis2015,Weis2017,Hormuth2021,Lorenzo2023}. 
Since the natural variations of ADC in healthy breast tissue may overlap with the ADC of areas of low tumor cell density, $N_{th}$ aims at delineating the simulated tumor region  that would be identified as a tumor from the MRI data (i.e., due to significantly low ADC values).
Hence, this threshold also enables the calculation of the tumor volume on model simulation results \textit{via} Eq.~\eqref{vt}.
Here, we set $N_{th}=\theta/4$, which includes the usual ADC values characterizing breast tumor regions in MRI datasets (e.g., see Fig.~\ref{fig_tboxes} and Refs. \cite{Jarrett2018,Jarrett2020,Jarrett2021}).
By calculating $N_T$ over the whole computational domain and $V_T$ over a thresholded region, we also assess whether there is a difference in evaluating model simulations using the whole model-calculated tumor cell density map $N(\mathbf{x},t)$ or just a restricted region.

For each parameter combination from matrices  $\mathbf{A}$, $\mathbf{B}$, and $\mathbf{A}_\mathbf{B}^{(i)}$, $i=1,\ldots,n_p$, we store the value of each model outcome of interest at time $t$ in vectors $\mathbf{Y}_\mathbf{A} (t)$, $\mathbf{Y}_\mathbf{B} (t)$, and $\mathbf{Y}_\mathbf{AB}^{(i)} (t)$, respectively, which each have length $n_s$.
Hence, the values in these vectors can correspond to either tumor cell volume $V_T$ or the total tumor cell count $N_T$.
Finally, the total-effects indices $S_{T,i} (t)$ for each parameter $p_i$, $i=1,\ldots,n_p$ on each model outcome of interest at each time step are calculated as

\begin{equation}\label{Sobol}
S_{T,i} (t) = \frac{1}{2 n_s \text{Var}\left(\mathbf{Y_A}(t), \mathbf{Y_B}(t)\right)}\sum_{j=1}^{n_s} \left( \left( \mathbf{Y_A}(t) \right)_j - \left( \mathbf{Y_{AB}}^{(i)}(t) \right)_j \right)^2
\end{equation}

\noindent where $ \text{Var}\left(\mathbf{Y_A}(t), \mathbf{Y_B}(t)\right)$ is the variance of all the values of the model outcome of interest at time $t$ obtained from all the computer simulations using all the parameter combinations in matrices $\mathbf{A}$ and $\mathbf{B}$.
If $S_{T,i} = 0$, then the parameter is non-influential and, hence, it can be fixed to any value within the range considered in the sensitivity analysis without affecting the variance of the model output of interest \cite{Saltelli2008}.
However, in practice it is common to set a tolerance $\epsilon_S$ such that if $S_{T,i} < \epsilon_S$, the parameter $p_i$ is considered non-influential on the model outcomes of interest.
Given that the patient-specific imaging data that is available to calibrate our model is scarce (see Section~\ref{intro}), we aim to reduce the number of parameters that will require personalized calibration as much as possible, so we set $\epsilon_S=0.1$ in the sensitivity analysis of this study.
In Section~\ref{resu}, we further compare the distribution of the model outcomes of interest (i.e., $V_T$ and $N_T$) obtained with the original model and the reduced model that results from fixing all non-influential parameters to a value within the range used in our sensitivity analysis (see Table~\ref{parspace}).
Additionally, 95$\%$ bootstrap confidence intervals for $S_{T,i}$ at any time $t$ can be calculated by resampling the values in $\mathbf{Y_A}(t)$, $\mathbf{Y_B}(t)$, and $\mathbf{Y}_\mathbf{AB}^{(i)} (t)$ in Eq.~\eqref{Sobol}.
In particular, we calculated these 95$\%$ bootstrap confidence intervals at the conclusion of NAC (i.e., at $t=T$) using the MATLAB function \textit{bootci} and 1000 bootstrap samples.

\subsubsection{Parameter values and ranges}\label{params}

Table~\ref{parspace} shows the parameter space considered in the sensitivity analysis for each NAC regimen.
The admissible values for the baseline tumor cell diffusivity $D_0$ and net proliferation rate $\rho_0$ were set based on our previous patient-specific, imaging-informed modeling studies on breast cancer response to NAC \cite{Weis2013,Weis2015,Weis2017,Jarrett2018,Jarrett2020,Jarrett2021}.
For each drug combination in the NAC regimens, the ranges for the Hill coefficients $h_i$, the half-maximal drug concentrations $C_i$, the synergy of potency coefficients $\alpha_i$ ($i=1,2$), and the synergy of efficacy coefficient $\beta$ were obtained from the \textit{in vitro} data described in Section~\ref{invitro}.
In particular, these experimentally-sourced ranges were defined to include all the 95$\%$ confidence intervals obtained for each parameter as a result of fitting the MuSyC equation to the \textit{in vitro} data collected for each TNBC line.
The intervals of admissible values for the normalized maximum concentration $\hat{d}_{m,i}$ and the decay rate $\gamma_i$ of the drugs in each NAC regimen ($i=1,2$) were constructed from our prior patient-specific, imaging-informed modeling studies \cite{Jarrett2018,Jarrett2020,Jarrett2021} and values reported in the literature \cite{Zoli1995,Zoli1999,Joerger2007,Joerger2007a,Anderson1996,Ohtsu1995,Vijgh1991,Jonge2005,Gianni1997,Kontny2013,Speth1988}.
Finally, the experimentally-determined values of the maximal drug effects $\hat{E}_i$ ($i=1,2$) for each NAC regimen only represented a reduction of tumor cell net proliferation or the induction on mild tumor cell death, which hindered the representation of the diverse NAC outcomes reported in clinical studies, ranging from tumor eradication (i.e., pCR) to progressive disease at NAC conclusion.
To avoid this issue, we estimated the corresponding clinically-relevant ranges for $\hat{E}_i$ ($i=1,2$) in each drug regimen \textit{in silico}.
To this end, we ran computer simulations with our model equipped with the rest of the parameter values within the aforementioned ranges.
These simulations sought to bound the value range of $\hat{E}_i$ ($i=1,2$) for each NAC regimen that rendered the diverse spectrum of therapeutic outcomes observed in clinical practice.
As observed in Table~\ref{parspace}, the admissible values for some of the parameters participating in the sensitivity analysis spanned several orders of magnitude.
In those cases, the parameter was sampled logarithmically, and the corresponding range in Table~\ref{parspace} corresponds to the decimal logarithm of the parameter.
For both the regularly and logarithmically sampled parameters, the distribution of admissible values was assumed to be uniform within the reported bounds in Table~\ref{parspace}.

The rest of the model parameters that were not included in the sensitivity analysis were fixed according to literature values and our prior patient-specific, imaging-informed modeling efforts \cite{Weis2013,Weis2015,Weis2017,Jarrett2018,Jarrett2020,Jarrett2021,Hormuth2021,Lorenzo2019,Lorenzo2020}.
We set the tumor cell carrying capacity $\theta=2.02\cdot 10^6$ cells/mm$^3$, the mechanical coupling coefficient $\gamma_N=2.5$ 1/KPa, the Winkler boundary constant $k_w=0.20$ KPa/mm, the tumor-induced pressure $g_N=5$ KPa, the Poisson's ratio $\nu=0.45$, and the Young modulus $E=3$ KPa.
While our model can accommodate heterogeneous mechanical properties that can be spatially defined \textit{via} MRI data \cite{Weis2013,Weis2015,Weis2017,Jarrett2018,Jarrett2020,Jarrett2021,Hormuth2021,Lorenzo2019,Lorenzo2020}, we opted for an average values over the breast tissue in an attempt to simplify and generalize the results of the sensitivity analysis beyond the particular background tissue architecture of the tumor scenarios considered in this study.

\section{Results}\label{resu}

\subsection{Doxorubicin and cyclophosphamide NAC regimen}

\begin{figure*}[!t]
\centering
\includegraphics[width=\linewidth]{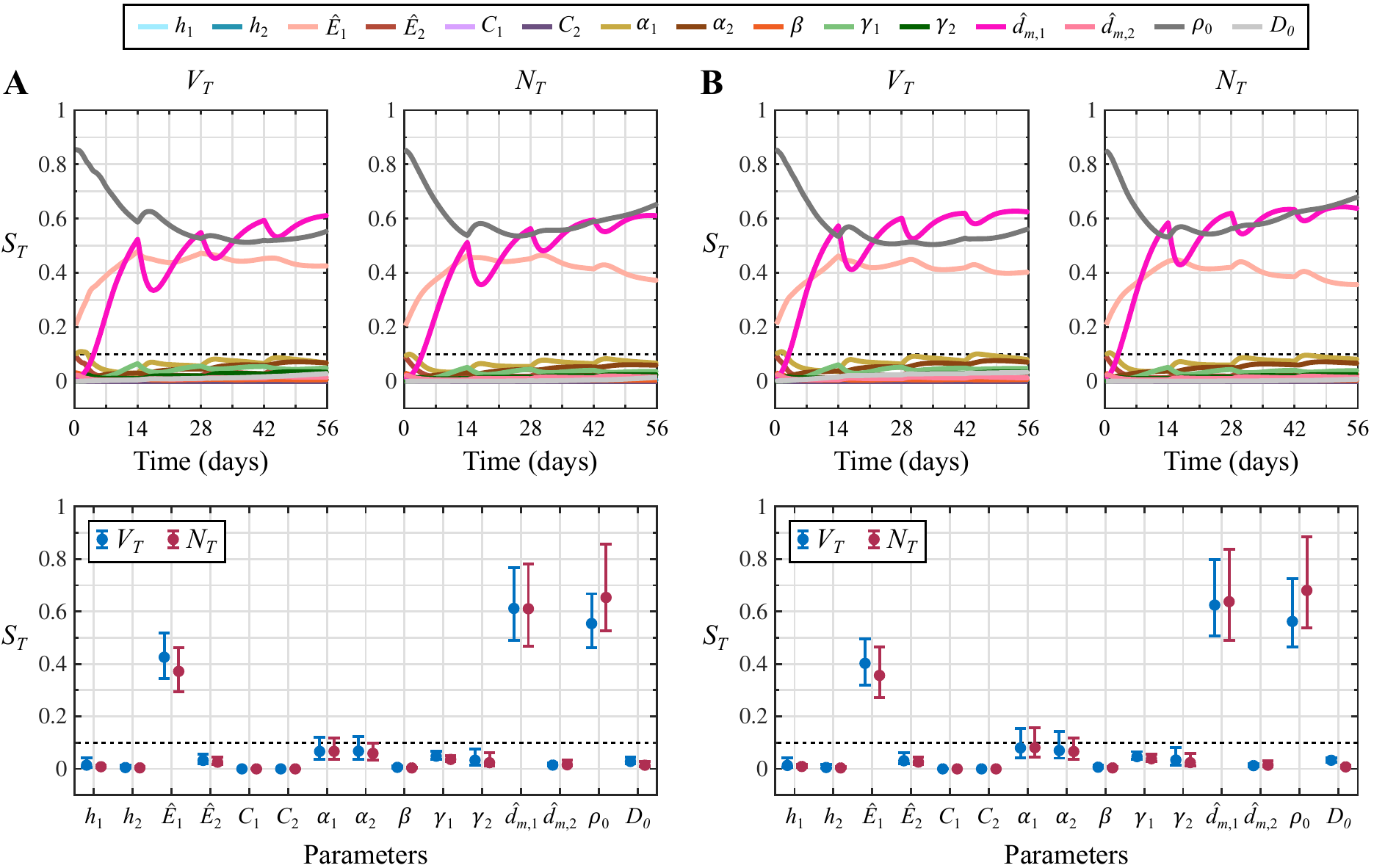}
\caption{\textbf{Sensitivity analysis results for the NAC regimen based on doxorubicin and cyclophosphamide.} 
Panel A shows the total effects indices ($S_T$) for each model parameter and quantity of interest ($V_T$ and $N_T$) obtained in the well-perfused tumor scenario.
Panel B provides the same results for the poorly-perfused tumor.
The top row in each panel depicts the dynamics of the total effects indices during the course of NAC.
The bottom row further shows the corresponding $S_T$ values at the conclusion of NAC along with their 95\% bootstrapped confidence interval.
Parameters with subindex 1 correspond to doxorubicin and parameters with subindex 2 are relative to cyclophosphamide.
The dashed line indicates the threshold of the total effects index to consider a parameter as influential ($S_T>0.1$) or non-influential ($S_T<0.1$).
Out of the $n_p=15$ model parameters considered in the sensitivity analysis, only three of them dominate the response of both tumors to the NAC regimen based on doxorubicin and cyclophosphamide: the baseline tumor cell net proliferation ($\rho_0$), the normalized maximal effect of doxorubicin ($\hat{E}_1$), and the normalized maximal drug concentration of doxorubicin ($\hat{d}_{m,1}$).
Additionally, these observations hold for the two quantities of interest considered in this study ($V_T$ and $N_T$), whose corresponding $S_T$ values for each parameter show similar dynamics and terminal values.
}
\label{fig_sadoxcyc}
\end{figure*}

Figure~\ref{fig_sadoxcyc} presents the results of the global variance-based sensitivity analysis corresponding to the NAC regimen based on doxorubicin and cyclophosphamide in the well-perfused and the poorly-perfused TNBC cases.
In both tumor scenarios, only three out of the $n_p=15$ model parameters included in the sensitivity analysis exhibit a total effects index on $V_T$ and $N_T$ over the threshold to identify influential parameters ($S_T>0.1$): the baseline tumor cell net proliferation ($\rho_0$), the normalized maximal effect of doxorubicin ($\hat{E}_1$), and the normalized maximal drug concentration of doxorubicin ($\hat{d}_{m,1}$).
This observation holds consistently both during the vast majority of the NAC regimen and at the conclusion of the treatment, when the corresponding 95\% confidence intervals of the total effects indices on $V_T$ and $N_T$ of the three influential parameters are also completely above the $S_T$ threshold.
As shown in Figure~\ref{fig_sadoxcyc}, while $\rho_0$ and $\hat{E}_1$ are identified as driving parameters since the onset of the NAC, $\hat{d}_{m,1}$ becomes influential during the first cycle.
Additionally, each cycle of this NAC regimen alters the rank of the $S_T$ values on $V_T$ and $N_T$ of the three influential parameters.
According to the global dynamics of the total effects indices on $V_T$ over the whole NAC regimen, $\rho_0$ is the most influential parameter at the beginning of treatment, but $\hat{d}_{m,1}$ progressively becomes the dominant model parameter towards the conclusion of NAC in both tumor scenarios.
The global dynamics of the total effects indices on $N_T$ show that $\rho_0$ is the most influential parameter over the majority of this NAC regimen and at the conclusion of the treatment in the two tumor cases, only momentarily superseded by $\hat{d}_{m,1}$ around the date of delivery of each NAC cycle.
Furthermore, following the delivery of the drug cycles at $t=14, 28,$ and $42$ days in both tumor scenarios, we observe that the values of the $S_T$ on $V_T$ and $N_T$ for parameter $\hat{d}_{m,1}$ drop while there is an increase of the corresponding $S_T$ values for parameters $\hat{E}_1$ and $\rho_0$.
However, the $S_T$ on $V_T$ and $N_T$ of parameter $\hat{d}_{m,1}$ then exhibits an increasing trend during each of these three last cycles, and may ultimately supersede the corresponding $S_T$ values for parameters $\hat{E}_1$ and $\rho_0$.
Among the non-influential parameters in both tumor scenarios,  the $S_T$ on $V_T$ and $N_T$ corresponding to the coefficients of synergy of potency of both doxorubicin and cyclophosphamide ($\alpha_1$ and $\alpha_2$, respectively) reach values close to $S_T=0.1$ according to the results provided in Figure~\ref{fig_sadoxcyc}.
Indeed, a fraction of their corresponding 95\% confidence interval at the conclusion of the NAC regimen is above this $S_T$ threshold.
Additionally, the  $S_T$ on $V_T$ and $N_T$ corresponding to the normalized maximal effect of cyclophosphamide ($\hat{E}_2$) and to the decay rates of doxorubicin and cyclophosphamide ($\gamma_1$ and $\gamma_2$)  also take values relatively close to the $S_T=0.1$ threshold according to the 95\% confidence intervals in both tumor scenarios.
Furthermore, this last observation also applies for the $S_T$ on $V_T$ of the baseline tumor cell diffusivity ($D_0$).

\begin{figure*}[!t]
\centering
\includegraphics[width=\linewidth]{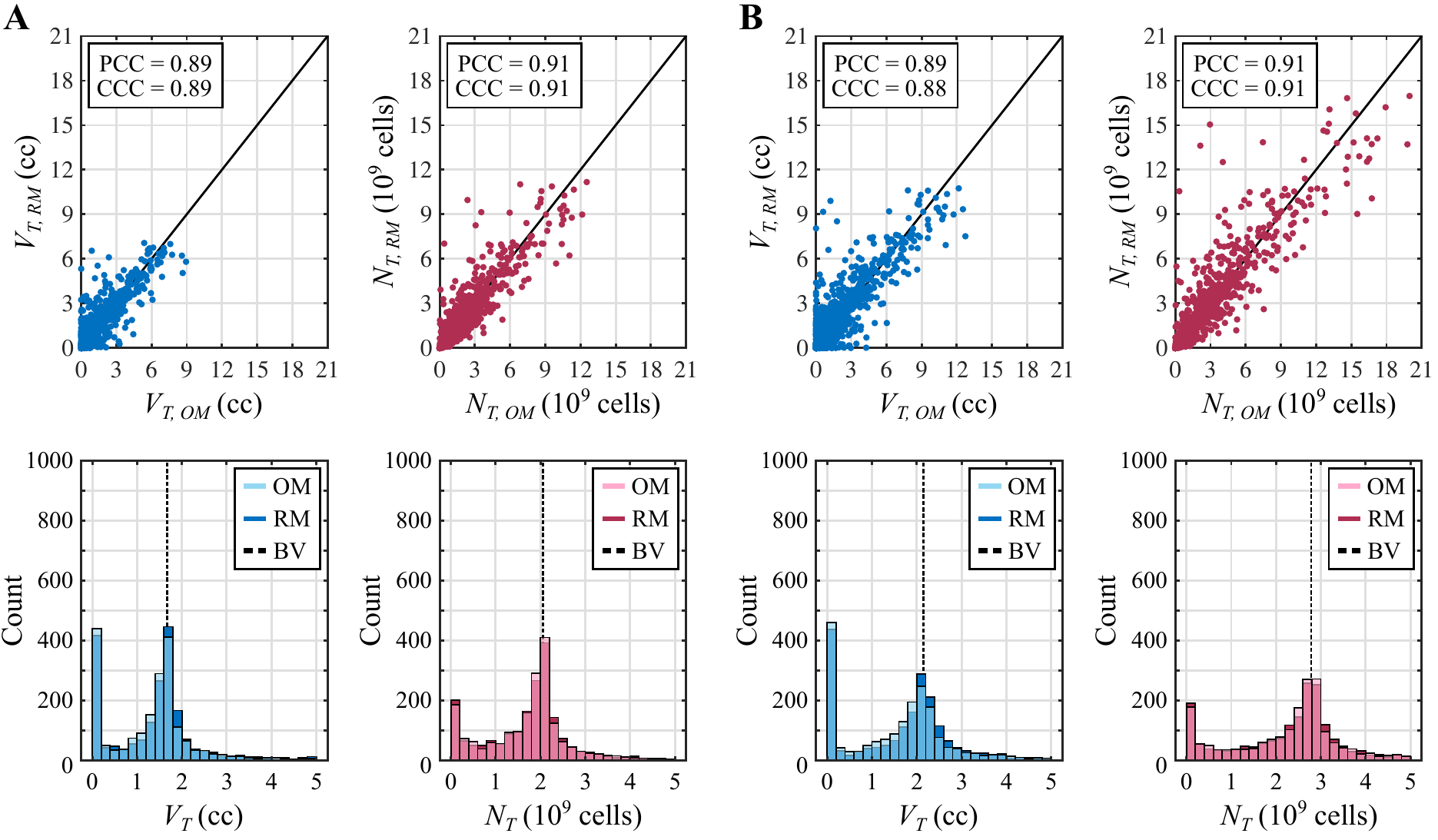}
\caption{\textbf{Comparison of the original and reduced models for the NAC regimen based on doxorubicin and cyclophosphamide.}  
Panel A compares the distribution of the tumor volume ($V_T$, blue) and the global tumor cell count ($N_T$, magenta) at the termination of NAC obtained using the original model (OM) and the reduced model (RM) in the well-perfused tumor. 
Panel B provides the corresponding results for the poorly-perfused tumor.
In each panel, the unity plots in the top row show the correlation between the $V_T$ and $N_T$ values obtained with either model ($n=2000$ for each quantity of interest and tumor scenario).
This correlation is further quantified in terms of the Pearson and the concordance correlation coefficients (PCC and CCC, respectively).
Additionally, the histograms in the bottom row of each panel compare the distribution of $V_T$ and $N_T$ values obtained with the original and reduced models in either tumor scenario.
A dashed vertical line indicates the baseline value (BV) of $V_T$ and $N_T$ at $t=0$ for each tumor, as a reference to assess NAC efficacy.
To facilitate the visualization of the main region of these distributions only the values of $V_T$ in $[0,5]$ cc and $N_T$ in $[0,5]\cdot 10^9$ cells are represented, which account for more than 90\% of the values obtained for each quantity of interest in either tumor (i.e., these plots do not depict higher values corresponding to outliers).
The plots depicted in this figure show good qualitative and quantitative agreement between the original and reduced models for the same choice of influential parameter values.
Importantly, these results further demonstrate that the solution space of both models contains a similar distribution of NAC outcomes, ranging from pCR to progressive disease.
}
\label{fig_comp_doxcyc}
\end{figure*}

\begin{figure*}[!t]
\centering
\includegraphics[width=\linewidth]{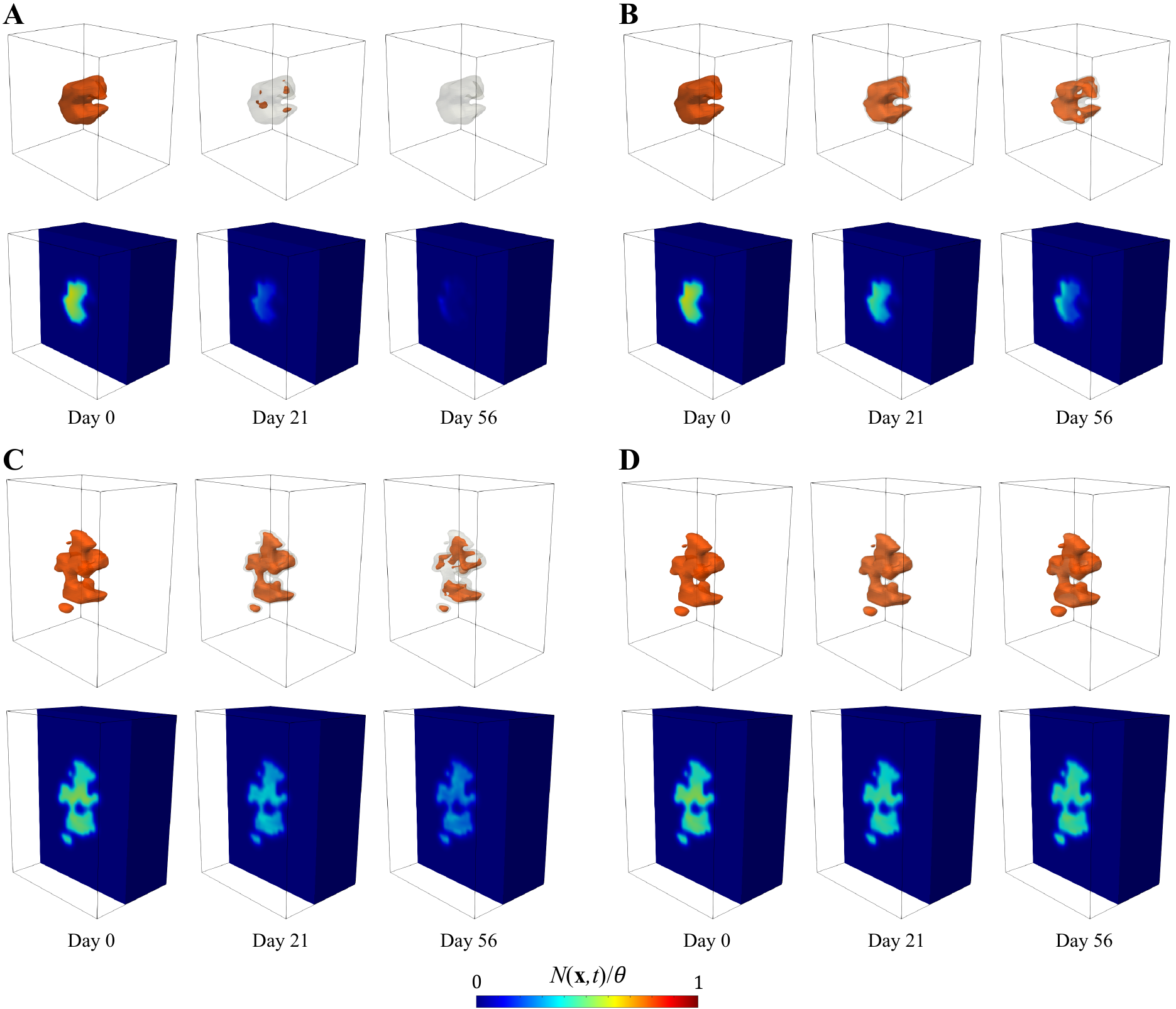}
\caption{\textbf{Examples of breast cancer response to the NAC regimen based on doxorubicin and cyclophosphamide.} 
Panels A and B illustrate two simulations of the response of the well-perfused tumor at the beginning (day 0), the middle of the second cycle (day 21), and the conclusion of NAC (day 56).
Likewise, panels C and D provide two examples of the response of the poorly-perfused tumor at the same three time points.
Each simulation corresponds to a choice of the model parameters within the bounds set in Table~\ref{parspace} (see Supplementary Table S5).
In each panel, the top row represents the outline of the tissue box including the evolving geometry of the tumor volume (orange) at the three time points, which is delineated over the model simulation results as $N(\mathbf{x},t)>N_{th}$ with $N_{th}=\theta/4$. 
These images further show the baseline tumor geometry at $t=0$ (gray volume) as a reference to assess the change in the tumor volume over the course of NAC.
The bottom row in each panel shows corresponding sagittal sections of the tumor cell density map $N(\mathbf{x},t)$.
The response scenario in panel A represents a pCR case, panels B and C provide mild responses to this NAC regimen, and panel D shows a non-responder scenario.
Notice that the status of the tumor by the middle of the second cycle is already informative of the expected therapeutic outcome at the conclusion of NAC.
We remark that this figure only provides two examples of response to the NAC regimen based on doxorubicin and cyclophosphamide for each of the two tumors considered in this study.
However, as shown in Fig.~\ref{fig_comp_doxcyc}, there are multiple model parameter combinations yielding responses to this NAC regimen that range from pCR to progressive disease for both tumors.}
\label{fig_simsdoxcyc}
\end{figure*}

The sensitivity analysis results for the NAC regimen based on doxorubicin and cyclophosphamide  justify the definition of a reduced model where only $\rho_0$, $\hat{E}_1$, and $\hat{d}_{m,1}$ need to be identified for each individual patient, while the rest of the model parameters can be fixed to any value within the ranges in Table~\ref{parspace}.
Figure~\ref{fig_comp_doxcyc} compares the distribution of $V_T$ and $N_T$ values obtained with the original and the reduced model at the conclusion of NAC in both tumor scenarios using the parameter combinations in matrices $\mathbf{A}$ and $\mathbf{B}$ that were leveraged during the sensitivity analysis ($n=2000$ parameter combinations; see Section~\ref{sa}).
Hence, for the reduced model, only the values corresponding to $\rho_0$, $\hat{E}_1$, and $\hat{d}_{m,1}$ are varied, while the rest of the parameters are fixed to the values reported in Supplementary Table S3 for all parameter combinations.
Figure~\ref{fig_comp_doxcyc} shows that the values of $V_T$ and $N_T$ obtained with either model at the end of the NAC regimen exhibit a high correlation in both tumor scenarios.
In particular, the Pearson and concordance correlation coefficients (PCC and CCC, respectively) for the $V_T$ values obtained with the original and the reduced model are both 0.89 in the well-perfused tumor, and 0.89 and 0.88 in the poorly-perfused tumor, respectively.
Similarly, the PCC and CCC of the $N_T$ values obtained with either model version are 0.91 in both tumor scenarios.
Figure~\ref{fig_comp_doxcyc} further provides a detail of the main part of the distribution (i.e., $>$90\% of values in the sample) of $V_T$ and $N_T$ values obtained with the original and the reduced model in the well-perfused and the poorly-perfused tumor.
These snapshots show that the distributions of $V_T$ and $N_T$ values obtained with either model are very similar.
Importantly, these distributions also show that both models can successfully recapitulate the diverse spectrum of TNBC responses to NAC based on doxorubicin and cyclophosphamide, ranging from a complete tumor eradication (i.e., pCR) to treatment failure exhibiting a larger tumor with respect to the corresponding baseline measurement (i.e., at $t=0$).
To further illustrate this diversity of TNBC responses to a doxorubicin and cyclophosphamide NAC regimen, Figure~\ref{fig_simsdoxcyc} shows four examples of the 3D dynamics of the tumors during this neoadjuvant regimen including a pCR, two mild responses, and treatment failure.
In particular, this figure shows both the change in the 3D volume of the tumor and the local map of tumor cell density (i.e., $N(\mathbf{x},t)$) halfway through the second NAC cycle ($t=21$ days) and at the end of treatment ($t=56$ days) with respect to  baseline ($t=0$).
We notice that, in the cases where there is a certain level of response (i.e, Figure~\ref{fig_simsdoxcyc}A-C), this phenomenon is more noticeable as a reduction of the values of the tumor cell density map than as a reduction of the tumor volumetric region. 
This observation supports the use of MRI measurements and computational predictions of tumor cell density maps $N(\mathbf{x},t)$ to characterize TNBC response to NAC rather than traditional tumor length or volume measurements.

\subsection{Paclitaxel and carboplatin NAC regimen}

\begin{figure*}[!t]
\centering
\includegraphics[width=\linewidth]{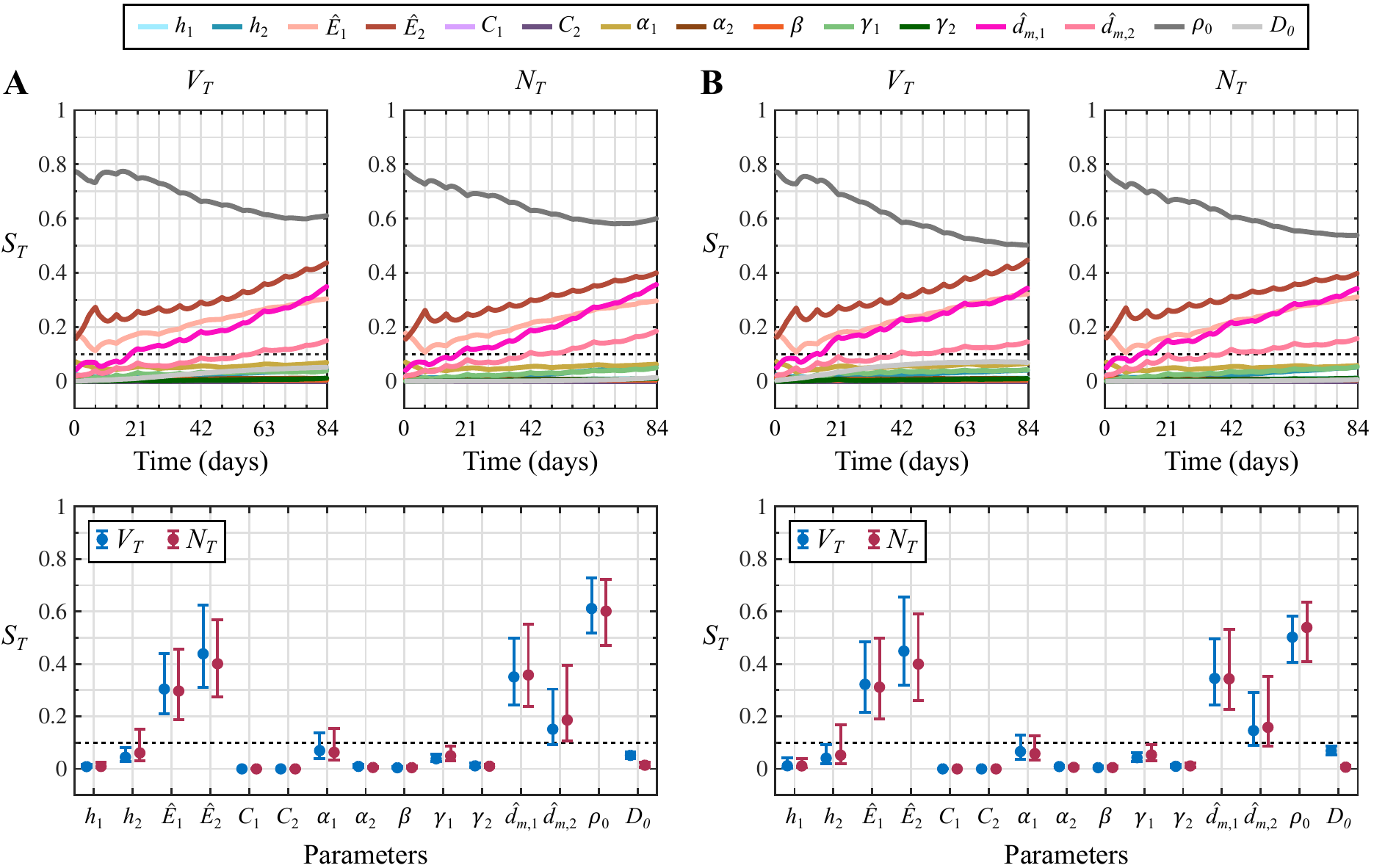}
\caption{\textbf{Sensitivity analysis results for the NAC regimen based on paclitaxel and carboplatin.}  
Panel A shows the total effects indices ($S_T$) for each model parameter and quantity of interest ($V_T$ and $N_T$) obtained in the well-perfused tumor scenario.
Panel B provides the same results for the poorly-perfused tumor.
The top row in each panel depicts the dynamics of the total effects indices during the course of NAC.
The bottom row further shows the corresponding $S_T$ values at the conclusion of NAC along with their 95\% bootstrapped confidence interval.
Parameters with subindex 1 correspond to paclitaxel and parameters with subindex 2 are relative to carboplatin.
The dashed line indicates the threshold of the total effects index to consider a parameter as influential ($S_T>0.1$) or non-influential ($S_T<0.1$).
Out of the $n_p=15$ model parameters considered in the sensitivity analysis, we observe that only five of them dominate the response of both tumors to the NAC regimen based on paclitaxel and carboplatin: the baseline tumor cell net proliferation ($\rho_0$), the normalized maximal effects of both paclitaxel and carboplatin ($\hat{E}_1$ and $\hat{E}_2$), and the normalized maximal drug concentration of both paclitaxel and carboplatin ($\hat{d}_{m,1}$ and $\hat{d}_{m,2}$).
Additionally, these observations hold for the two quantities of interest considered in this study ($V_T$ and $N_T$), whose corresponding $S_T$ values for each parameter show similar dynamics and terminal values.}
\label{fig_saptxcpt}
\end{figure*}

Figure~\ref{fig_saptxcpt} provides the results of the global variance-based sensitivity analysis corresponding to the NAC regimen based on paclitaxel and carboplatin in the well-perfused and the poorly-perfused TNBC scenarios.
In both cases, only five out the $n_p=15$ model parameters considered in the sensitivity analysis reach values of the total effects index on $V_T$ and $N_T$ over the threshold to identify influential parameters ($S_T>0.1$): the baseline tumor cell net proliferation ($\rho_0$), the normalized maximal effects of both paclitaxel and carboplatin ($\hat{E}_1$ and $\hat{E}_2$), and the normalized maximal drug concentration of both paclitaxel and carboplatin ($\hat{d}_{m,1}$ and $\hat{d}_{m,2}$).
However, not all these parameters are classified as influential during the whole course of the paclitaxel and carboplatin NAC regimen.
While $\rho_0$, $\hat{E}_1$, and $\hat{E}_2$ exhibit $S_T>0.1$ on both $V_T$ and $N_T$ from the beginning of the NAC regimen, the corresponding $S_T$ values for $\hat{d}_{m,1}$ only become influential by the end of the first NAC cycle and for $\hat{d}_{m,2}$ by the end of the second or third cycle (see Figure~\ref{fig_saptxcpt}).
At the conclusion of the NAC regimen, the five parameters are classified as influential and their corresponding 95\% confidence intervals are above the $S_T=0.1$ threshold, except for a minimal fraction for  $\hat{d}_{m,1}$ and $\hat{d}_{m,2}$.
Regarding the rank of the driving parameters, we observe that $\rho_0$ and $\hat{E}_2$ are consistently identified as the first and second most influential parameters according to the total effects indices on $V_T$ and $N_T$ in both tumor scenarios, while $\hat{d}_{m,2}$ always  shows the lowest $S_T$ values among the five influential parameters detected for this NAC regimen.
Additionally, while $\hat{E}_1$ is more influential than $\hat{d}_{m,1}$ at the beginning of treatment, their rank is reversed towards the conclusion of the NAC regimen.
This last observation also holds both for the total effects indices on $V_T$ and $N_T$ as well as for both tumor scenarios.
Figure~\ref{fig_saptxcpt} further shows that the weekly delivery of paclitaxel introduces local peaks or minima in the values of the total effects indices on $V_T$ and $N_T$, but that the overall dynamics during the whole course of this NAC regimen further account for the action of carboplatin (see Figure~\ref{fig_pkpd}).
Among the non-influential parameters in both tumor scenarios, the $S_T$ of $V_T$ and $N_T$ of the Hill coefficient of carboplatin ($h_2$), the decay rate of paclitaxel ($\gamma_1$), and the coefficient of synergy of potency of paclitaxel ($\alpha_1$) are close to the $S_T$ threshold by the end of treatment, and a fraction of their 95\% confidence interval may even  be above the $S_T=0.1$ line (see Figure~\ref{fig_sadoxcyc}).
Additionally, the $S_T$ on $V_T$ of the baseline tumor cell diffusivity ($D_0$) also reaches values near the $S_T=0.1$ threshold in both tumor scenarios.

\begin{figure*}[!t]
\centering
\includegraphics[width=\linewidth]{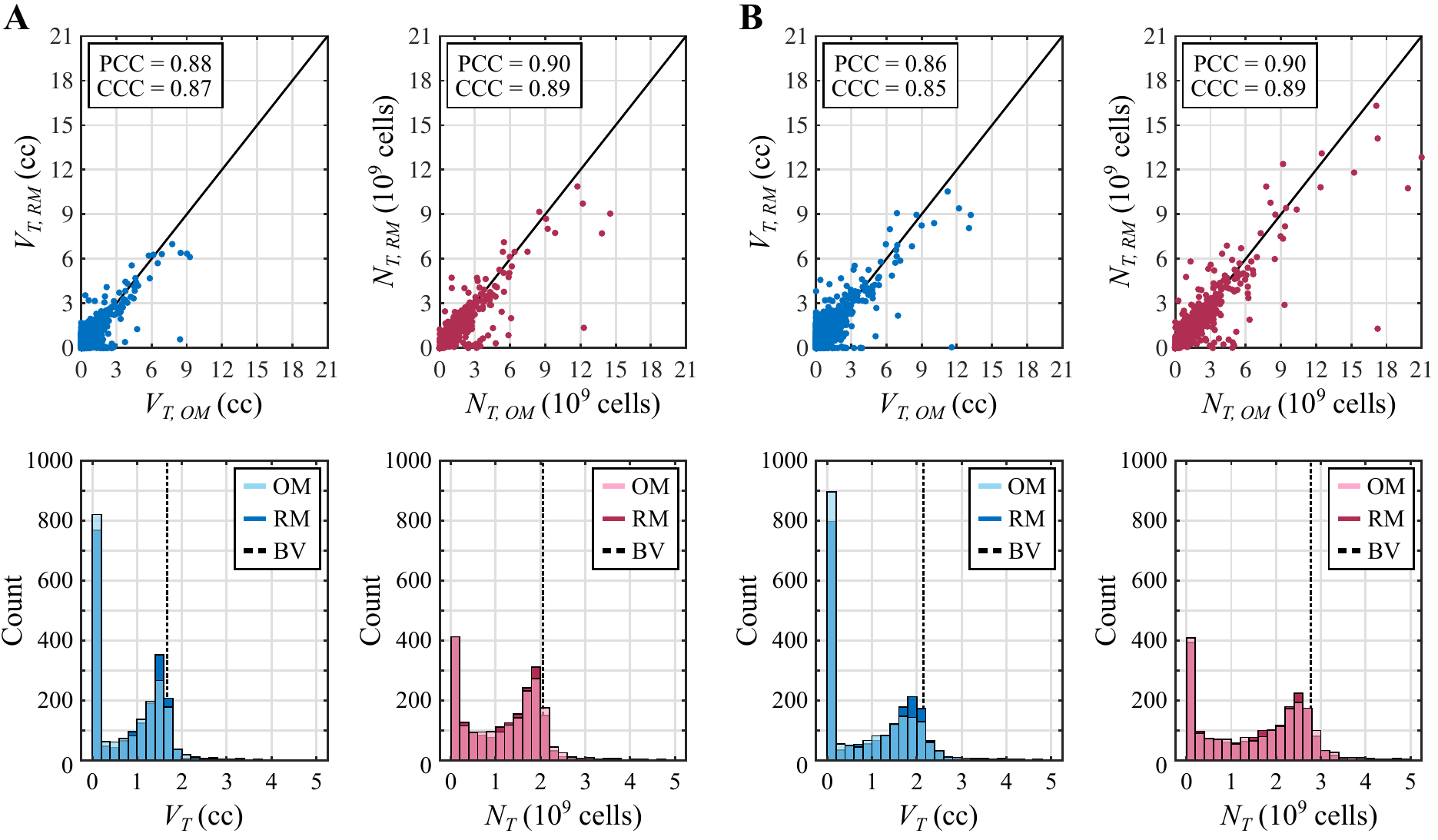}
\caption{\textbf{Comparison of the original and reduced models for the NAC regimen based on paclitaxel and carboplatin.}  
Panel A compares the distribution of the tumor volume ($V_T$, blue) and the global tumor cell count ($N_T$, magenta) at the termination of NAC obtained using the original model (OM) and the reduced model (RM) in the well-perfused tumor. 
Panel B provides the corresponding results for the poorly-perfused tumor.
In each panel, the unity plots in the top row show the correlation between the $V_T$ and $N_T$ values obtained with either model ($n=2000$ for each quantity of interest and tumor scenario).
This correlation is further quantified in terms of the Pearson and the concordance correlation coefficients (PCC and CCC, respectively).
Additionally, the histograms in the bottom row of each panel compare the distribution of $V_T$ and $N_T$ values obtained with the original and reduced models in either tumor scenario.
A dashed vertical line indicates the baseline value (BV) of $V_T$ and $N_T$ at $t=0$ for each tumor, as a reference to assess NAC efficacy.
To facilitate the visualization of the main region of these distributions only the values of $V_T$ in $[0,5]$ cc and $N_T$ in $[0,5]\cdot 10^9$ cells are represented, which account for more than 90\% of the values obtained for each quantity of interest in either tumor (i.e., these plots do not depict higher values corresponding to outliers).
The plots depicted in this figure show good qualitative and quantitative agreement between the original and reduced models for the same choice of influential parameter values.
Importantly, these results further demonstrate that the solution space of both models contains a similar distribution of NAC outcomes, ranging from pCR to progressive disease.}
\label{fig_comp_ptxcpt}
\end{figure*}

\begin{figure*}[!t]
\centering
\includegraphics[width=\linewidth]{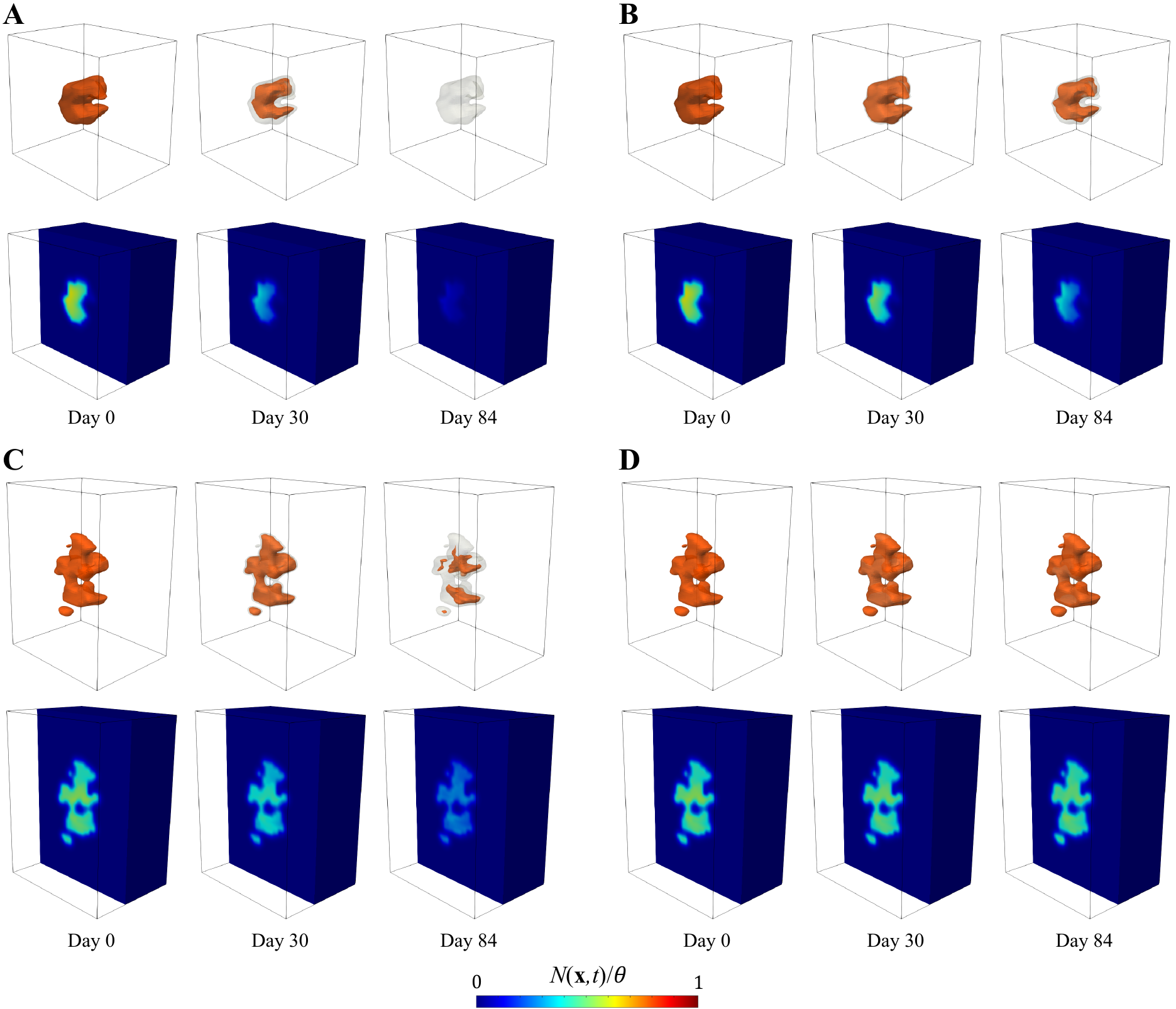}
\caption{\textbf{Examples of breast cancer response to the NAC regimen based on paclitaxel and carboplatin.}  
Panels A and B illustrate two simulations of the response of the well-perfused tumor at the beginning (day 0), the middle of the second cycle (day 30), and the conclusion of NAC (day 84).
Likewise, panels C and D provide two examples of the response of the poorly-perfused tumor at the same three time points.
Each simulation corresponds to a choice of the model parameters within the bounds set in Table~\ref{parspace} (see Supplementary Table S6).
In each panel, the top row represents the outline of the tissue box including the evolving geometry of the tumor volume (orange) at the three time points, which is delineated over the model simulation results as $N(\mathbf{x},t)>N_{th}$ with $N_{th}=\theta/4$. 
These images further show the baseline tumor geometry at $t=0$ (gray volume) as a reference to assess the change in the tumor volume over the course of NAC.
The bottom row in each panel shows corresponding sagittal sections of the tumor cell density map $N(\mathbf{x},t)$.
The response scenario in panel A represents a pCR case, panels B and C provide mild responses to this NAC regimen, and panel D shows a non-responder scenario.
Notice that the status of the tumor by the middle of the second cycle is already informative of the expected therapeutic outcome at the conclusion of NAC.
We remark that this figure only provides two examples of response to the NAC regimen based on paclitaxel and carboplatin for each of the two tumors considered in this study.
However, as shown in Fig.~\ref{fig_comp_ptxcpt}, there are multiple model parameter combinations yielding responses to this NAC regimen that range from pCR to progressive disease for both tumors.}
\label{fig_simsptxcpt}
\end{figure*}

Based on the results of the sensitivity analysis for the paclitaxel and carboplatin NAC regimen, we can define a reduced model where only $\rho_0$, $\hat{E}_1$, $\hat{E}_2$, $\hat{d}_{m,1}$, and $\hat{d}_{m,2}$ require patient-specific identification, while the rest of the model parameters can be fixed to any value within the ranges defined in Table~\ref{parspace}.
Figure~\ref{fig_comp_ptxcpt} provides a comparison between the distribution of $V_T$ and $N_T$ values obtained with the original and the reduced model at the termination of this NAC regimen in the well-perfused and the poorly-perfused tumor scenario.
Again, we leveraged the parameter combinations in matrices $\mathbf{A}$ and $\mathbf{B}$ of the sensitivity analysis ($n=2000$; see Section~\ref{sa}).
Hence, for the reduced model we only vary the values of $\rho_0$, $\hat{E}_1$, $\hat{E}_2$, $\hat{d}_{m,1}$, while the rest of the parameters are fixed to the values provided in Supplementary Table S4.
According to the unity plots presented in Figure~\ref{fig_comp_ptxcpt}, the values of $V_T$ and $N_T$ obtained with either model at the conclusion of NAC are highly correlated in both tumor scenarios.
In particular, the PCC and CCC for the $V_T$ values obtained with the original and the reduced model in the well-perfused tumor are 0.88 and 0.87, respectively, while in the poorly-perfused tumor the corresponding PCC and CCC values are 0.86 and 0.85, respectively.
Likewise, the PCC and CCC of the $N_T$ values obtained with either model version are 0.90 and 0.89, respectively, in both the well-perfused and the poorly-perfused scenarios.
Figure~\ref{fig_comp_ptxcpt} also shows  a detail of the main part of the distribution (i.e., $>$90\% of values in the sample) of the $V_T$ and $N_T$ values obtained with the original and the reduced model in each tumor scenario.
We observe that these distributions exhibit good qualitative agreement, demonstrating that our modeling framework can successfully reproduce the different responses of TNBC to an NAC regimen based on paclitaxel and carboplatin, ranging from the complete elimination of the tumor to progressive disease by the end of treatment.
Furthermore, Figure~\ref{fig_simsptxcpt} provides illustrative examples of the 3D dynamics of TNBC response to this NAC regimen, including a pCR, two mild responses, and a treatment failure.
In particular, this figure depicts both the 3D tumor region and the local map of tumor cell density (i.e., $N(\mathbf{x},t)$) at baseline ($t=0$ days), halfway during the second NAC cycle ($t=30$ days), and at the conclusion of treatment ($t=84$ days).
As for doxorubicin and cyclophosphamide, we observe that a positive TNBC response to the paclitaxel and carboplatin NAC regimen (e.g., see Figure~\ref{fig_simsptxcpt}A-C) is more noticeable as a reduction of the local values of $N(\mathbf{x},t)$ rather than a decrease in the tumor region. Hence, this observation also supports the use of MRI measurements and computational forecasts of tumor cell density maps to assess TNBC response to NAC over conventional metrics relying on tumor length or volume.

\section{Discussion}\label{discussion}
NAC has been widely regarded as an opportunity to optimize the management of TNBC for individual patients by adjusting their treatment based on the early tumor response to the initially prescribed drug regimen \cite{Liu2010,Barker2009,Minckwitz2013,Gupta2020,Jarrett2020,Jarrett2021}.
This approach could specifically introduce a major improvement in the management of TNBC, since this breast cancer subtype lacks targeted therapies, tends to exhibit an adverse prognosis, and is hard to treat successfully.
To characterize NAC response early in the course of treatment, quantitative measurements of tumor cellularity and perfusion obtained \textit{via} DW-MRI and DCE-MRI, respectively, have shown a superior performance  with respect to standard metrics relying on geometric features of the tumor (e.g., longest axis, volume) \cite{Eisenhauer2009,Hylton2012,Tudorica2016,Pickles2006}.
The integration of these imaging measurements from an individual patient within mechanistic models of breast cancer growth has further shown promise not only in recapitulating the spatiotemporal dynamics of the patient's tumor response, but also in predicting the corresponding therapeutic outcome at the conclusion of NAC \textit{via} personalized computer simulations  \cite{Weis2013,Weis2015,Jarrett2018,Jarrett2020,Jarrett2021}.
In this context, this work presents a mechanistic model that describes the spatiotemporal dynamics of breast response to NAC.
Our model characterizes the baseline tumor growth in terms of tumor cell density using an established formulation consisting of a combination of net proliferation, \textit{via} a logistic term, and tumor cell mobility, \textit{via} a mechanically-constrained diffusion operator \cite{Weis2013,Weis2015,Jarrett2018,Jarrett2020,Jarrett2021,Lorenzo2023}.
The mechanical deformation induced by the tumor is modeled as a quasistatic process driven by a growth term depending on the local tumor cell density within the linear elastic constitutive equation of breast tissue.
Additionally, our model captures the tumor response to NAC by adjusting the baseline tumor cell net proliferation rate using a term that integrates the NAC drug pharmacokinetics and pharmacodynamics, including potential drug-drug synergies according to the MuSyC framework \cite{Meyer2019}.

In this work, we performed a global variance-based sensitivity analysis to identify the dominant parameters  that represent the model mechanisms driving breast cancer response to two usual NAC regimens: doxorubicin plus cyclophosphamide, and paclitaxel plus carboplatin \cite{Waks2019,Gupta2020,Jarrett2020,Poggio2018}.
Towards this end, we constrained the parameter space of our model by integrating  \textit{in vitro} measurements of the response of TNBC cell lines to the two aforementioned NAC drug combinations and \textit{in silico} estimates of TNBC dynamics during these NAC regimens from  personalized  forecasts informed by \textit{in vivo} quantitative MRI data.
Since the NAC drugs considered in this study are delivered intravenously, we further perform the sensitivity analysis in two different scenarios corresponding to a well-perfused and a poorly-perfused tumor.
These two tumor scenarios were constructed by leveraging  quantitative MRI data from two TNBC patients, respectively.
The resulting values of the total effects indices ($S_T$) on tumor volume ($V_T$) and total tumor cell count ($N_T$) show that only a minority of the model parameters considered in the sensitivity analysis ($n_p=15$) dominate the response of the tumor cases to the two NAC regimens investigated in this study ($S_T>0.1$): three in the case of doxorubicin plus cyclophosphamide, and five in the case of paclitaxel plus carboplatin (see Figures~\ref{fig_sadoxcyc} and \ref{fig_saptxcpt}).
The baseline tumor cell proliferation rate ($\rho_0$) was always identified as an influential parameter, and in the paclitaxel and carboplatin NAC regimen it represents the primary mechanism driving the tumor response over the whole course of the treatment.
Additionally, our sensitivity analysis also identified the normalized maximal effects ($\hat{E}$) and the normalized maximal drug concentration ($\hat{d}_{m}$) of doxorubicin, paclitaxel, and carboplatin as dominant parameters.
No model parameter related to cyclophosphamide was identified as relevant in this study (see Figure~\ref{fig_sadoxcyc}).
This might be due to the relatively fast wash-out rate of this drug compared to doxorubicin \cite{Jonge2005,Speth1988,Kontny2013,Anderson1996}.
Hence, our sensitivity analysis results suggest that it suffices to fix the parameters characterizing the pharmacodynamics and pharmacokinetics of cyclophosphamide within the ranges of Table~\ref{parspace}.
Taking together the dominant model parameters, our sensitivity analysis concludes that drug-induced changes of the baseline tumor cell net proliferation constitute the dominant mechanism characterizing TNBC response to the NAC regimens investigated herein.
Our results further indicate that these changes are ultimately driven by the maximum drug concentration reaching the tumor and the maximum drug-induced cytotoxic effect on the baseline tumor cell net proliferation.
Additionally, our results  support the use of longitudinal MRI measurements of tumor cellurarity obtained \textit{via} DW-MRI to assess tumor response to NAC, given that drug-induced changes of tumor cell net proliferation result in local variations of the tumor cell density according to our model (see Section~\ref{tumdym}) \cite{Weis2013,Weis2015,Jarrett2018,Jarrett2020,Jarrett2021,Lorenzo2023}.
Indeed, the 3D simulations provided in Figures~\ref{fig_simsdoxcyc} and \ref{fig_simsptxcpt} show that the NAC-induced reduction in tumor cell density renders more insight into the therapeutic response of TNBC than the change of the tumor volume region.

To assess the ability of the dominant model parameters in driving TNBC response to NAC, we further compared the distribution of $V_T$ and $N_T$ values at NAC conclusion obtained with our model for the parameter combinations used in the sensitivity analysis ($n=2000$) to the corresponding distributions of these quantities of interest obtained with a reduced version of the model, in which only the dominant parameters are varied while the non-influential parameters are fixed to values within the ranges in Table~\ref{parspace} (see Supplementary Tables S3 and S4).
This analysis showed a high correlation between the $V_T$ values obtained with either model, which reached a PCC$>$0.86 and CCC$>0.85$, as well as between the corresponding $N_T$ values, for which PCC$>$0.90 and CCC$>0.89$, across both tumor scenarios and NAC regimens (see Figures~\ref{fig_comp_doxcyc} and \ref{fig_comp_ptxcpt}).
Additionally, Figures~\ref{fig_comp_doxcyc}, \ref{fig_simsdoxcyc}, \ref{fig_comp_ptxcpt}, and \ref{fig_simsptxcpt} show that both versions of the model produce a similar distribution of $V_T$ and $N_T$ values, ranging from tumor elimination (i.e., pCR) to progressive disease, where the tumor burden is larger than at the onset of NAC.
These results demonstrate that the parameter space constructed in this study by integrating \textit{in vivo}, \textit{in vitro}, and \textit{in silico} data of TNBC response to NAC contains parameter combinations that enable both the original and reduced models to span tumor responses across the diverse spectrum observed in clinical practice \cite{Minckwitz2012,Yau2022,Eisenhauer2009,Hylton2012,Jarrett2020}.
Thus, the quantitative and qualitative agreement between the original and the reduced model support the use of the reduced model as a surrogate to characterize the dynamics of TNBC response to NAC in individual patients.
Importantly, the reduced parameter set requiring patient-specific identification ($n_p=3$ or 5, depending on the NAC regimen, \textit{versus} $n_p=15$ in the original model) constitutes a practical advantage to deploy our predictive modeling framework in the neoadjuvant setting due to the scarcity of longitudinal MRI data on therapeutic response for each individual patient during the course of NAC, which can only reliably inform a limited number of model parameters.

From a modeling perspective, the choice of the values for the non-influential parameters can introduce an important simplification in the model formulation.
For both drugs in the two NAC regimens in this study, the Hill coefficients ($h_1$ and $h_2$) and the coefficients of drug synergy of potency ($\alpha_1$ and $\alpha_2$) are classified as non-influential in the sensitivity analysis ($S_T<0.1$).
If we select the values of these four parameters within the ranges in Table~\ref{parspace} such that $\alpha_2^{h_1}=\alpha_1^{h_2}$, then $g_0\left(\hat{d}_1,\hat{d}_2\right)=g_1\left(\hat{d}_1,\hat{d}_2\right)=g_2\left(\hat{d}_1,\hat{d}_2\right)=g_3\left(\hat{d}_1,\hat{d}_2\right)/\alpha_1^{h_2}$ in Eqs.~\eqref{g0}-\eqref{g3} and, hence, Eq.~\eqref{msyc} can be simplified to

\begin{equation}\label{msyceasy}
\rho = \rho_0 \frac{ 1 + \hat{d}_1^{h_1}\hat{E}_1 + \hat{d}_2^{h_2}\hat{E}_2 + \alpha_1^{h_2}\hat{d}_1^{h_1}\hat{d}_2^{h_2}\hat{E}_3}{ 1 + \hat{d}_1^{h_1} + \hat{d}_2^{h_2} + \alpha_1^{h_2}\hat{d}_1^{h_1}\hat{d}_2^{h_2} },
\end{equation}

\noindent where $\hat{E}_3$ is provided by Eq.~\eqref{e3}.
The relationship  $\alpha_2^{h_1}=\alpha_1^{h_2}$ implies that the drugs are in detailed balance \cite{Meyer2019,Wooten2021}, which ultimately assumes that the cytotoxic effect of the drugs in each NAC regimen is instantaneous.
Importantly, the sensitivity analysis results from our work support the adoption of this assumption for TNBC forecasting during NAC regimens based on either doxorubicin and cyclophosphamide or paclitaxel and carboplatin.
Hence, the simpler formulation of Eq.~\eqref{msyceasy} facilitates the mechanistic analysis and code implementation of our mechanistic model of TNBC response to NAC.

While the sensitivity analysis and ensuing modeling study presented in this work shows promise for the use of our modeling framework to render personalized tumor forecasts of TNBC response to NAC, it also features several limitations.
First, in this work we focused on TNBC and two NAC regimens, but our modeling framework can also be extended to other breast cancer subtypes and neoadjuvant drugs \cite{Zardavas2015,Waks2019,Gupta2020,Spring2022}.
To this end, our multiscale framework would need representative tumor scenarios using \textit{in vivo} MRI data characterizing the tumor subtypes.
It would also be necessary to define an appropriate parameter space by obtaining  \textit{in vitro} measurements of tumor cell response to the neoadjuvant drugs of interest for each breast cancer subtype.
Additionally, the parameter space could also be refined by integrating \textit{in silico} estimates of model parameters from tumor forecasts for patients exhibiting each subtype and receiving the neoadjuvant drugs under investigation. 
Second, the well-perfused and poorly-perfused tumor scenarios on the tissues boxes leveraged in this study provide a practical framework to analyze the dynamics of breast cancer response to NAC using our mechanistic model. However, further studies should reassess the ability of the reduced model to recapitulate TNBC dynamics during NAC in patient-specific breast geometries and accounting for diverse tumor perfusion maps \cite{Jarrett2018,Jarrett2020,Jarrett2021}.
Third, we assumed that the tumor perfusion map did not change over time in our work. 
This is a common assumption in prior tumor forecasting studies between consecutive measurements of tumor perfusion \textit{via} DCE-MRI \cite{Jarrett2018,Jarrett2020}.
Nevertheless, our model could be extended to account for the dynamics of tumor-supported vasculature (i.e., angiogenesis) during NAC \cite{Hormuth2021a,Lorenzo2023,Hormuth2020,Xu2017,Travasso2011,Swanson2011,Kohandel2007,Kremheller2019}, which could also be the target of therapeutic regimens involving antiangiogenic drugs (e.g., bevacizumab) \cite{Zardavas2015,Gupta2020,Bear2012,Colli2021,Kohandel2007}.
This vasculature-coupled model could be further extended by including a computational fluid dynamics representation of blood flow and drug delivery over the breast anatomy \cite{Hormuth2021a,Wu2020,Wu2022,Fritz2021,Kremheller2019,Vavourakis2018}.
This extension of the model could contribute to achieve a more precise quantification of the drug concentrations in the tumor and healthy tissue, thereby enabling a finer assessment of NAC efficacy and toxicity \cite{Wu2022}.
Fourth, we leveraged linear elasticity to characterize the TNBC-induced tissue deformation and the effect of local mechanical stress on tumor cell mobility, which has been shown to be a reasonable approximation during the time course of usual NAC regimens \cite{Weis2013,Weis2015,Weis2017}.
Future studies could investigate a mechanical coupling with tumor cell proliferation, which can also be inhibited by local mechanical stress \cite{Lorenzo2019,Lima2017,Stylianopoulos2013,Jain2014,Fraldi2018} and might thus reduce the efficacy of chemotherapeutic agents \cite{Rizzuti2020}.
Moreover, a poroelastic formulation of breast tissue would refine the quantification of solid stress and interstitial pressure effects on tumor dynamics and drug distribution, which could render a more accurate description of  the spatiotemporal dynamics of local drug availability and treatment effects \cite{Stylianopoulos2013,Fraldi2018,Kremheller2019}.
Fifth, we considered that the model parameters were globally defined over the tumor region.
This assumption simplifies the analysis of our model but offers a limited representation of the heterogeneous morphology and ensuing NAC response dynamics that TNBC may exhibit.
Alternatively, intratumoral heterogeneity could be captured by leveraging local, spatially-resolved parameter maps \cite{Weis2013,Weis2015,Jarrett2018,Jarrett2020,Jarrett2021} or by defining parameter subsets characterizing the dynamics of distinct tumor habitats featuring unique cellularity and perfusion profiles identified \textit{via} quantitative MRI imaging (e.g., DW-MRI and DCE-MRI) \cite{Kazerouni2022}.
Sixth, we adopted a deterministic framework in the sensitivity analysis and modeling study performed in this work.
Future studies could extend our analysis by leveraging a Bayesian framework, which provides a robust methodology to further account for the uncertainties in the different sources of the parameter estimates as well as in the model outcomes (e.g., $V_T$ and $N_T$) \cite{Lima2017,Kapteyn2021,Lorenzo2023,Lipkova2019}.
Moreover, a Bayesian framework would allow for investigating alternative model versions within a model selection study \cite{Lima2017,Lorenzo2023}.
These alternative models may include all or a subset of the dominant parameters identified in our sensitivity analysis, recruit some non-influential parameters to boost model performance, and feature some of the extensions of the model formulation discussed above.
Finally, the computer simulations in our work rely on isogeometric analysis, which is a recent generalization of classic finite element analysis \cite{Cottrell2009}.
This is an established approach to perform a moderate number of deterministic simulations of the model, for example, during the calibration of global parameters and ensuing tumor forecasting for an individual patient \cite{Lorenzo2023}.
However, standard finite-element and isogeometric methods can become computationally intensive for some of the aforementioned model extensions.
The computational cost can become prohibitive for a Bayesian implementation for uncertainty quantification and model selection, since these techniques require a large number of model simulations to construct probability distributions \cite{Lima2017,Lorenzo2023,Kapteyn2021}.
To address these limitations, reduced order modeling approaches can be used to accelerate computer simulations of tumor growth models  \cite{Viguerie2022,Chinesta2016}.

In the future, we plan to investigate the performance of the reduced model identified in this study on recapitulating TNBC response during NAC using \textit{in vivo} quantitative MRI data of individual patients in a diverse cohort (e.g., different tumor locations in the breast, sizes, and perfusion profiles).
We further plan to assess whether model forecasts of TNBC response obtained by informing the model with just two longitudinal MRI datasets acquired early in the course of NAC (e.g., at baseline and after one or two cycles) suffice to accurately predict therapeutic outcome at the conclusion of the prescribed NAC regimen for each patient \cite{Jarrett2018,Jarrett2020,Jarrett2021}.
This capability would enable the treating oncologist to identify non-responding patients early in the course of NAC, who could benefit from switching to other therapeutics, as well as exceptional responders, who could benefit from de-escalation or even avoidance of the ensuing breast surgery \cite{Barker2009,Minckwitz2013,Veeraraghavan2017,Spring2022,Gupta2022,Heil2020}.
Additionally, while clinical studies cannot evaluate the complex landscape of potential drug combinations, dosages, and scheduling strategies (e.g., timing and order of drug delivery), a validated mechanistic model of TNBC response to NAC supported by a robust parameter space characterizing potential baseline tumor dynamics and combined drug effects would constitute a powerful technology to rigorously and systematically design optimal therapeutic plans for individual patients \cite{Jarrett2020,Colli2021,Wu2022,Lorenzo2023}.
Hence, this technology could be cast as a digital twin of the patient's tumor that assimilates upcoming imaging or clinical data as they become available to update the prediction of NAC outcome \cite{Wu2022a,Wu2022b,Kapteyn2021,Niederer2021}.
Then, this digital twin would provide the treating oncologist with a comprehensive picture of observed and expected tumor dynamics during the prescribed NAC regimen and potential alternative treatments, thereby supporting optimal clinical decision-making for each individual patient.

\section{Conclusion}
We have presented a new mechanistic model which extends an established mechanically-coupled, drug-informed formulation of breast cancer response to NAC with an explicit term representing the pharmacokinetics and pharmacodynamics of the drugs in the NAC regimen.
We further investigate the mechanisms in the model that drive TNBC response to two usual NAC regimens (doxorubicin plus cyclophosphamide, and paclitaxel plus carboplatin) by performing a global variance-based sensitivity analysis in two distinct scenarios: a well-perfused and poorly-perfused tumor, which are defined upon two corresponding patient-specific \textit{in vivo} MRI data.
To this end, we constructed the parameter space of our model by integrating  \textit{in vitro} measurements of TNBC cell response to the NAC drugs of interest \textit{via} high-throughput automated microscopy, and \textit{in silico} estimates of model parameters from prior MRI-informed tumor forecasting studies in TNBC patients receiving the NAC regimens under investigation.
Our sensitivity analysis identified the drug-induced changes in tumor cell net proliferation caused by doxorubicin, paclitaxel, and carboplatin as the main drivers of TNBC response to the two NAC regimens considered herein.
Out of the $n_p=15$ parameters included in the sensitivity analysis, our results show that these driving mechanisms are ultimately regulated by only three parameters in the doxorubicin and cyclophosphamide regimen (the baseline tumor cell net proliferation, the normalized maximal effect of doxorubicin, and the normalized maximal concentration of doxorubicin) and only five parameters in the paclitaxel and carboplatin regimen (the baseline tumor cell net proliferation, the normalized maximal effect of both drugs, and the normalized maximal concentration of both drugs).
Additionally, we demonstrated that a reduced version of our model that is solely driven by the dominant parameters can be leveraged as a valid surrogate of our original model, since both models achieve a high level of qualitative and quantitative agreement while also enabling the simulation of the diverse spectrum of NAC outcomes observed in the clinic (i.e, from pCR to progressive disease).
Thus, we posit that the reduced model is an excellent candidate to compute personalized MRI-informed predictions of NAC outcome, and we further envision that this model could ultimately support a personalized digital twin to optimize clinical decision-making during NAC for each individual TNBC patient.

\section*{Acknowledgments}
G.L. was partially supported by a Peter O'Donnell Jr. Postdoctoral Fellowship from the Oden Institute for Computational Engineering and Sciences at The University of Texas at Austin and acknowledges funding from the European Union’s Horizon 2020 research and innovation programme under the Marie Sk\l{}odowska-Curie grant agreement No. 838786. 
We thank the National Institutes of Health for funding through NCI U01CA142565, U01CA174706, R01CA186193, U24CA226110, R50CA243783, and U01CA253540. 
We also thank the Cancer Prevention and Research Institute of Texas for support through CPRIT RR160005; T.E.Y is a CPRIT Scholar in Cancer Research. 
We further acknowledge the Texas Advanced Computing Center (TACC) for providing high-performance computing resources that contributed to the results presented in this work.

\section*{Author contributions}

{
\setlength{\parskip}{0pt}
Conceptualization: G.L., D.R.T., V.Q., T.E.Y. 

Methodology: G.L., A.M.J., C.T.M., D.R.T., V.Q., T.E.Y.

Software: G.L., D.R.T. 

Validation: G.L., D.R.T.

Formal Analysis: G.L., D.R.T., T.E.Y.

Investigation: G.L., D.R.T., V.Q., T.E.Y.

Resources: G.L., A.M.J., C.T.M., D.R.T., V.Q., T.E.Y.

Data Curation: G.L., A.M.J., C.T.M., D.R.T.

Writing - Original Draft Preparation: G.L., A.M.J., D.R.T., T.E.Y.

Writing - Review \& Editing:  G.L., A.M.J., C.T.M., D.R.T., V.Q., T.E.Y.

Visualization: G.L., D.R.T. 

Supervision: G.L., T.E.Y. 

Project Administration: G.L., T.E.Y.

Funding Acquisition: D.R.T., V.Q., T.E.Y.
}

\section*{Declaration of interests}
The authors declare no conflict of interest.


\begin{small}

\end{small}


\end{document}